\documentclass[12pt, a4paper, fleqn]{article}

\usepackage[nottoc]{tocbibind} 
\usepackage{multicol}

\RequirePackage[l2tabu, orthodox]{nag}
\usepackage{silence}
\ActivateWarningFilters[pdftoc]

\usepackage[T1]{fontenc}
\usepackage[utf8]{inputenc}

\usepackage[top=19mm, bottom=19mm, left=25mm, right=25mm]{geometry}

\usepackage{amsmath,amssymb,amsfonts,amscd,latexsym,amsthm}

\newtheoremstyle{break}{9pt}{9pt}{\itshape}{}{\bfseries}{}{\newline}{}
\theoremstyle{break}
\newtheorem{defn}[equation]{Definition}
\newtheorem{conj}[equation]{Conjecture}
\newtheorem{prop}[equation]{Proposition}

\newtheorem{ax}[equation]{Axiom}

\usepackage{xcolor}  % Otherwise 'Undefined color: gray' in BBL file.

\usepackage{pgf}
\usepackage{tikz}
\usetikzlibrary{shapes.misc}
\tikzset{cross/.style={cross out, draw=black, thick, fill=none, minimum size=2*(#1-\pgflinewidth), inner sep=0pt, outer sep=0pt}, cross/.default={3pt}}

\usepackage[colorlinks=true, linktoc=all, linkcolor=black, citecolor=red, urlcolor=blue, backref=page]{hyperref}

\usepackage{etoolbox}
\makeatletter
\patchcmd{\BR@backref}{\newblock}{\newblock(}{}{}
\patchcmd{\BR@backref}{\par}{)\par}{}{}
\makeatother

\numberwithin{equation}{section}
\usepackage{float}

\begin{document}

\

\begin{flushleft}
{\bfseries\sffamily\Large 
Conformal correlator systems % with abelian monodromies
\vspace{1.5cm}
\\
\hrule height .6mm
}
\vspace{1.5cm}

{\bfseries\sffamily 
Sylvain Ribault
}
\vspace{3mm}

{\textit{\ \ 
Institut de physique théorique, CEA, CNRS, Université Paris-Saclay %, 91191, Gif-sur-Yvette, France
}}
\vspace{2mm}

{\textit{\ \ E-mail:}\texttt{
sylvain.ribault@ipht.fr
}}
\end{flushleft}
\vspace{7mm}

{\noindent\textsc{Abstract:}
In critical limits of statistical models, there exist non-local correlators that are conformally covariant but do not admit operator product expansions. To describe such correlators, we define conformal correlator systems, whose axioms are weaker than those of conformal field theory. As a result, $N$-point correlators are determined by $4$-point correlators, instead of $3$-point correlators in CFT.

In two dimensions, conformal spins may take arbitrary complex values, leading to correlators with nontrivial monodromies. We show that sphere $4$-point correlators and torus $1$-point correlators obey nontrivial monodromy constraints. We construct correlators with abelian monodromies in free bosonic theories and in critical loop models.
}

\clearpage

\hrule 
\tableofcontents
\vspace{5mm}
\hrule
\vspace{5mm}

\hypersetup{linkcolor=blue}

\section{Introduction and summary}\label{sec:intro}

\subsubsection{Critical lattice models with or without OPEs}

In statistical physics, for any $Q\in\mathbb{N}_{\geq 2}$, the $Q$-state Potts model describes the dynamics of spins $\sigma^a$ with $a=1,\dots ,Q$ on a $d$-dimensional lattice. This generalizes the Ising model, which is the case $Q=2$. For some values of $(d,Q)$, the model has a critical limit \cite{wj23}. In this limit, the spins $\sigma^a$ give rise to primary fields, whose correlators are described by a conformal field theory.
In particular, these correlators satisfy the fundamental axiom of CFT: they admit \textbf{operator product expansions}. Namely, two spins that are close enough can be replaced with a linear combination of fields, \textit{whose coefficients do not depend on the rest of the correlator}.

The $Q$-state Potts model can be generalized to any $Q\in\mathbb{C}$, by trading the spins $\sigma^a$ for clusters, and spin correlators for cluster connectivities \cite{fk72}. Since clusters are non-local variables, there is no reason for OPEs to survive this generalization. For integer $Q$, cluster connectivities may not admit OPEs, but at least they are linear combinations of spin correlators \cite{dv11}. For non-integer $Q$, the spins $\sigma^a$ no longer exist, while we still have conformal symmetry in the critical limit. To what conformally invariant structure do cluster connectivities belong, if not a CFT?

This question is sharper in the case $d=2$, where we have some analytic control over connectivities of the critical $Q$-state Potts model, and over more general correlators of critical loop models, see \cite{rib24} for a review.
In particular, there is strong evidence that these correlators are linear combinations of conformal blocks. This may look like a consequence of conformal symmetry, since conformal blocks form a basis of conformally covariant functions. However, the blocks that appear form a small subset of that basis. And the coefficients of the linear combinations do not factorize into products of $3$-point structure constants, as we would expect in CFT as a consequence of OPEs.

\subsubsection{Generalizing OPEs and CFTs}

To account for correlators that are conformally covariant but do not admit OPEs, we will propose generalizations of the notions of OPE and CFT:
\begin{align}
 \begin{tikzpicture}[baseline=(base), scale = .5]
 \coordinate (base) at (0, 0);
 \node at (-8, 1) {$\boxed{\text{operator product expansion}}$};
 \node at (-8, -1) {$\boxed{\text{conformal field theory}}$};
 \node at (8, 1) {$\boxed{\text{correlator expansion}}$};
 \node at (8, -1) {$\boxed{\text{conformal correlator system}}$};
 \draw[red, ultra thick, -latex] (-1.2, 1) -- (2.8, 1);
  \draw[red, ultra thick, -latex] (-2.4, -1) -- (1.6, -1);
 \end{tikzpicture}
\end{align}
A correlator expansion (Axiom \ref{ax:ce}) is defined like an OPE, with however coefficients that depend on the whole correlator, and not just on the two involved fields. In this sense, correlator expansions are non-local, while OPEs are local.
A conformal correlator system (Definition \ref{def:ccs}) is a set of correlators that admit correlator expansions instead of OPEs.

In CFT, by performing multiple OPEs, we can reduce $N$-point correlators to $3$-point correlators. In conformal correlator systems, we can reduce $N$-point correlators to $4$-point correlators instead. This follows from crossing symmetry equations, which involve changes of bases of conformal blocks. Any change of basis can be decomposed into changes of bases of $4$-point blocks, therefore any crossing symmetry equation can be reduced to the $4$-point case, and its solutions can be expressed in terms of $4$-point correlators.

\subsubsection{Two-dimensional correlators, single-valued or multivalued}

In two dimensions, we will not only weaken OPEs, but also allow the fields' conformal spins to take arbitrary complex values $S\in\mathbb{C}$. Let us argue that it is possible to relax the known constraints on $S$:
\begin{itemize}
 \item Assuming that \textbf{sphere correlators are single-valued} implies $S\in\frac12\mathbb{Z}$. But in 2d, moving a point $z_1$ around $z_2$ is topologically nontrivial. We can therefore consider multivalued correlators, with nontrivial monodromies. We will say that a correlator has \textbf{abelian monodromies} if it only picks up constant prefactors under such moves, and \textbf{non-abelian monodromies} otherwise. (This terminology is inspired by abelian and non-abelian anyons.)
 \item Assuming that \textbf{the torus partition function exists and is modular invariant} implies $S\in\mathbb{Z}$, where the torus partition function is defined as a trace over the space of states. However, the torus partition function is a feature of the operator formalism, and needs not exist in the bootstrap approach. What does exist is a torus $0$-point correlator, which is modular invariant, but is in general a sum over only part of the spectrum. States that do not contribute to the torus $0$-point correlator need not have integer spins.
\end{itemize}
In the case of critical loop models, allowing complex spins is not only possible, but in fact necessary. (See Section \ref{sec:clm}.)

\subsubsection{Multivalued correlators and their expansions}

To describe monodromies and modular transformations of primary fields, we introduce their \textbf{phases} $\theta\in\mathbb{C}^*$:

\begin{figure}[H]
\centering
 \begin{tabular}{|c|c|c|}
 \hline
  Name & Notation & Definition
  \\
  \hline \hline
  Left and right conformal dimensions & $\Delta,\bar\Delta$ & Eigenvalues of  $L_0,\bar L_0$
  \\
  \hline
  Conformal spin & $S$  & $S=\Delta-\bar \Delta$
  \\
  \hline
  Phase & $\theta$ & $\theta = e^{2\pi i S}$
  \\
  \hline
 \end{tabular}
 \refstepcounter{equation}
 \caption{Features of primary fields}
\end{figure}
Consider an expansion of two primary fields into a linear combination of primary and descendant fields:
\begin{align}
 V_1(z_1)V_2(z_2) = \sum_k C_k \left|z_{12}^{\Delta_k-\Delta_1-\Delta_2}\right|^2 \Big(V_k(z_2) + \text{descendants}\Big)\ ,
\end{align}
where we use the notation $\left|z^\delta\right|^2 = z^\delta\bar z^{\bar \delta}$ (where $\bar z$ is the complex conjugate of $z\in\mathbb{C}$). If $C_k$ only depends on $V_1,V_2,V_k$ this is an OPE; if it depends on the correlator where our two fields are inserted this is a correlator expansion. Now, if $z_1$ moves around $z_2$, let us write the monodromy of this expansion:
\begin{align}
 V_1(z_1)V_2(z_2)\ \underset{z_{12}\to e^{2\pi i}z_{12}}{\longrightarrow}\  \sum_k C_k \left(\theta_k\theta_1^{-1}\theta_2^{-1}\right)\left|z_{12}^{\Delta_k-\Delta_1-\Delta_2}\right|^2 \Big(V_k(z_2) + \text{descendants}\Big)\ ,
 \label{tk12}
\end{align}
where the factor $\big(V_k(z_2) + \text{descendants}\big)$ is unchanged, since it is a power series in $z_{12},\bar z_{12}$. This monodromy is trivial if $\theta_k =\theta_1\theta_2$, abelian if $\theta_k$ is $k$-independent, and non-abelian otherwise.

A sphere $N$-point correlator $\left<V_1(z_1)\cdots V_N(z_N)\right>$ belongs to a representation of the fundamental group of the sphere minus $N-1$ points, acting via the monodromies of $z_1$ around $z_2,\dots, z_N$. Together with global conformal symmetry, this leads to constraints that are nontrivial for $N\geq 4$. For a $4$-point correlator $\left<V_1V_2V_3V_4\right>$ with abelian monodromies, let $\theta_s,\theta_t,\theta_u$ be the phases of the fields that appear in the expansions $V_1V_2,V_1V_4,V_1V_3$ respectively, then we will find the constraint (Section \ref{sec:mono})
\begin{align}
 \boxed{\theta_s\theta_t\theta_u = \theta_1\theta_2\theta_3\theta_4} \ .
 \label{tstttu}
\end{align}
For a torus $1$-point correlator $\left<V_1\right>^{(1)}$, the analogous constraint is (Section \ref{sec:1pt})
\begin{align}
 \boxed{\theta^{12} = \theta_1}\ ,
 \label{tt1}
\end{align}
where $\theta$ is the phase of channel fields, $\theta_1$ the phase of $V_1$, and $12$ is the number twelve.

\subsubsection{Examples}

We now restrict to the two-dimensional case, and we assume local conformal symmetry: this gives us access to correlators that can be studied analytically. We classify our examples according to the existence of OPEs, and to the monodromies of correlators:
\begin{table}[H]
\centering
 \begin{tabular}{r|c|c|}
   & \textcolor{green!60!black}{With OPEs} & \textcolor{green!60!black}{No OPEs}
   \\
   \hline
   \textcolor{green!60!black}{Single-valued} & &
   Critical loop models
   \\
   \hline
   \textcolor{green!60!black}{Abelian} & Free bosonic correlators & Deformed critical loop models
   \\
   \hline
   \textcolor{green!60!black}{Non-abelian} &
   \begin{tabular}{c} Coulomb gas integrals \\ Critical XXZ$_q$ spin chains \end{tabular}
   & \begin{tabular}{c} Extended critical loop models \\ Topological defects \end{tabular}
   \\
   \hline
 \end{tabular}
 \refstepcounter{equation}
 \caption{Examples of conformal correlator systems}
\end{table}
Let us briefly introduce these conformal correlator systems, before elaborating in Section \ref{sec:ex}:
\begin{itemize}
 \item \textbf{Free bosonic correlators} are a family of conformal correlators, which admit OPEs \cite{rib14}. These include correlators in some CFTs, but no CFT contains all free bosonic correlators (for a given central charge). We obtain a multivalued generalization by decoupling the left-moving and right-moving momentums. Momentum conservation implies that monodromies are abelian.
  \item \textbf{Coulomb gas integrals} are a family of conformal correlators, which generalize free bosonic correlators, and admit OPEs. These integrals include correlators in some CFTs, such as minimal models. Multivalued generalizations exist, with monodromies that need not be abelian. For example, the \textbf{critical XXZ$_q$ spin chains} of Gabai et al \cite{ggqzz24} involve non-integer spins, and their correlators are Coulomb gas integrals with non-abelian monodromies.
 \item \textbf{Critical loop models}, including the critical $Q$-state Potts model, give rise to single-valued correlators that admit no OPEs. Describing these correlators is our initial motivation for introducing conformal correlator systems.
 Moreover, we will propose multivalued generalizations:
 \begin{itemize}
 \item A \textbf{deformed critical loop model} allows non-integer spins, while requiring monodromies to be abelian. We call it deformed, because any correlator tends to a correlator of a critical loop model, when spins tend to integer values.
 We will numerically solve crossing symmetry for sphere $4$-point correlators, and find just as many solutions as in the single-valued case, provided the consistency constraint \eqref{tstttu} is obeyed.
 \item An \textbf{extended critical loop model} is a non-abelian generalization, where phases take finitely many values in each channel. Then crossing symmetry equations have more solutions than in the single-valued case.
 \end{itemize}
 \item Consider an $N$-point correlator of some CFT (say a minimal model), in the presence of closed \textbf{topological defects} of codimension one. A defect that separates two fields modifies the correlator expansion of these fields, which is no longer an OPE. Moreover, in the case $d=2$, if a field moves around another field while dragging a defect along (and not crossing it), the topology of the defect changes, therefore single-valuedness is lost:

 \begin{figure}[H]
\centering
  \begin{tikzpicture}[baseline=(base), scale = .5]
\coordinate (base) at (0, 1.3);
\draw (0, 0) node[cross]{};
  \draw (3, 0) node [cross]{};
  \draw (0, 3) node[cross]{};
  \draw (3, 3) node[cross]{};
  \node at (0, 2.3) {$z_2$};
  \node at (3, 2.3) {$z_3$};
  \node at (0, -.8) {$z_1$};
  \node at (3, -.7) {$z_4$};
   \draw[red, thick] (0, -.4) to [out = 0, in = 0] (0, 3.4) to [out = 180, in = 180] (0, -.4);
  \end{tikzpicture}
 $ \qquad\quad
  \underset{z_{23}\to e^{2\pi i}z_{23}}{\longrightarrow}
  \qquad\quad$
  \begin{tikzpicture}[baseline=(base), scale = .5]
\coordinate (base) at (0, 1.3);
\draw (0, 0) node[cross]{};
  \draw (3, 0) node [cross]{};
  \draw (0, 3) node[cross]{};
  \draw (3, 3) node[cross]{};
  \draw[red, thick] (-.3, -.3) to [out = -45, in = -45] (4, 4) to [out = 135, in = 90] (-.4, 3) to [out = -90, in = 135] (3.3, 3.3) to [out = -45, in = 135] (-.3, -.3);
  \end{tikzpicture}
 \refstepcounter{equation}
 \caption{Behaviour of a closed topological defect under monodromy}
 \label{defect}
 \end{figure}
 There are two possible interpretations of our correlator:
 \begin{align}
 \boxed{ \begin{tabular}{l}
   $N$ fields plus defects
   \\ in a CFT
  \end{tabular}}
  \ \ \simeq \ \
  \boxed{\begin{tabular}{l}
  $N$ fields
  \\
  in a conformal correlator system
  \end{tabular}}
 \end{align}
\end{itemize}

\section{Conformal correlator systems on the sphere}

In this section we give the definition and properties of conformal correlator systems on the $d$-dimensional Euclidean sphere. In particular, we introduce the axiom of correlator expansions, which is weaker than the existence of OPEs, but still implies that correlators decompose into conformal blocks.

The behaviour of correlators under conformal transformations is the same as in CFT, see the review \cite{prv18}. We will focus on the properties of structure constants: the coefficients of the decompositions into conformal blocks. Our formulas are expected to hold in any dimension, although our examples will be in $d=2$.

\subsection{Spectrums and correlators}

In this section, we define our basic objects: spectrums, fields, and correlators.

\begin{defn}[Spectrum]
~\label{spec}
A spectrum is a set $\mathcal{S}$ of indecomposable representations of
 the conformal algebra.
\end{defn}

For simplicity, we assume that each representation $R$ appears at most once, in other words its multiplicity obeys $m_R\in \{0,1\}$. Allowing nontrivial multiplicities $m_R\in \mathbb{N}\cup\{\infty\}$ would make $\mathcal{S}$ a multiset.

\begin{defn}[Field]
 ~\label{field}
 Given a spectrum, a field $V_{R,v}$ is a representation $R\in\mathcal{S}$, together with a vector $v\in R$. We may alternatively use the notation $V_i$, then $R_i,v_i$ are the corresponding representation and vector.
\end{defn}

A field is nothing more than a convenient notation for writing properties of correlators.
If all representations with $R\in\mathcal{S}$ are highest-weight representations, the spectrum may be identified with the set of primary fields.

\begin{defn}[Correlator]
~\label{def:cor}
Given $N$ fields in a spectrum, an $N$-point correlator $\big<V_1(x_1)\cdots V_N(x_N)\big>_p$ on the $d$-dimensional Euclidean sphere $\mathbb{S}$ is a complex function on
\begin{align}
\mathbb{X}_N = \left\{(x_1,\cdots ,x_N)\in \mathbb{S}^N\middle| i\neq j\implies x_i\neq x_j \right\} \ ,
\label{xn}
\end{align}
which depends linearly on each one of the vectors $v_1,\cdots ,v_N$, and may depend on extra parameters $p$. For $\pi\in S_N$ a permutation of $\{1,2,\dots, N\}$, we assume $\big<V_1(x_1)\cdots V_N(x_N)\big>_p=\big<V_{\pi(1)}(x_{\pi(1)})\cdots V_{\pi(N)}(x_{\pi(N)})\big>_p$.
\end{defn}

In the example of topological defects, parameters of the defects (including their topology) are extra parameters. In the example of connectivities of the $Q$-state Potts model, there is a discrete extra parameter: the partition of $\{1,2,\cdots, N\}$ that describes whether the points $x_1,x_2,\cdots , x_N$ belong to the same of different clusters.

\begin{defn}[Basis]
 A basis $\mathcal{B}=(B_R)_{R\in\mathcal{S}}$ of the spectrum $\mathcal{S}$ is a collection of bases $B_R$ for each representation, such that every element of each basis is a generalized eigenvector of the dilation operator. (Its generalized eigenvalues are called conformal dimensions.) For $\mathcal{B}$ a basis, the corresponding set of fields is $(V_{R,v})_{R\in\mathcal{S},v\in B_R}$.
\end{defn}

The precise definition of a basis depends on the analytic properties of the spectrum. In unitary CFTs, representations are Hilbert spaces, and we are dealing with Hilbert bases. More generally, we expect representations to be Banach spaces, in which case we have Schauder bases. By definition, the space of finite linear combinations of elements of a Schauder basis is dense in the Banach space.

\subsection{Correlator expansions}

\begin{ax}[Correlator expansion]
~\label{ax:ce}
For $N\geq 2$, any $N$-point correlator has a $V_iV_j$ expansion, for any $1\leq i<j\leq N$. In the case of a $V_1V_2$ expansion, this means that our correlator is a linear combination of $(N-1)$-point correlators of the type
\begin{align}
 \left<V_1(x_1)V_2(x_2)\cdots \right>_p = \sum_{R,v\in\mathcal{B}} \sum_q C_R^{p, q} f_{R,v}(x_1,x_2) \left<V_{R,v}(x_2)\cdots\right>_q\ ,
 \label{vovt}
\end{align}
such that the sum converges on a neigbourhood of $\{x_1=x_2\}$.
The coefficients factorize into constants $C_R^{p, q}$, and universal factors $f_{R,v}(x_1,x_2)$ that depend neither on $V_3,\cdots,V_N$, nor on the extra parameters $p,q$.
\end{ax}
The universal factors $f_{R,v}$ are determined by conformal symmetry (plus a normalization condition), so they are the same as in OPEs.
In particular, if $V_1,V_2,V_{R,v}$ are scalar primary fields, then
$f_{R,v}(x_1,x_2) = |x_1-x_2|^{\Delta_R-\Delta_1-\Delta_2}$, where $\Delta$ is the conformal dimension. For simplicity we assume $\dim\operatorname{Hom}(R_1\otimes R_2,R)\leq 1$, otherwise $C_R^{p, q}$ and $f_{R,v}(x_1,x_2)$ would depend on an extra index that would distinguish different tensor structures.

\begin{defn}[Operator product expansion]
Two fields $V_1,V_2$ admit an OPE if the constants $C_R^{p,q}$ depend only on $V_1$ and $V_2$, and neither on $V_3,\cdots , V_N$ nor on $p,q$.
 \end{defn}

\begin{defn}[Conformal correlator system]
~\label{def:ccs}
Given a spectrum with a basis $\mathcal{B}$, a conformal correlator system is a set of functions $\left<V_{R_1,v_1}V_{R_2,v_2}\cdots V_{R_N,v_N}\right>$ on $\mathbb{X}_N$ for any $R_i,v_i\in \mathcal{B}$, such that:
\begin{itemize}
 \item The correlators transform covariantly under the conformal algebra, according to the representations $R_i$ and vectors $v_i$. In other words, they obey conformal Ward identities.
 \item The correlators admit expansions $V_{R_i,v_i}V_{R_j,v_j}$ on open subsets $\Omega_{i,j}\subset \mathbb{X}_N$, such that $\Omega_{i,j}\cap \Omega_{k,\ell}\neq \emptyset$.
\end{itemize}
\end{defn}

\begin{prop}[Crossing symmetry]
 Any correlator can be decomposed into conformal blocks, with coefficients called structure constants that do not depend on the fields' positions. These coefficients are constrained by crossing symmetry: the equality between different decompositions of the same correlator.
\end{prop}

This proposition follows from the existence of correlator expansions, in the same way as it follows from the existence of operator product expansions. For this purpose it does not matter which variables $C^{p,q}_R$ depends on, only that it is constant. Therefore, the definition of conformal blocks and the statement of crossing symmetry are the same as in CFT \cite{prv18}.

\begin{defn}[CFT]
 A CFT is a conformal correlator system such that all correlator expansions are OPEs.
\end{defn}

\subsection{Equations for structure constants}

For a $4$-point correlator, crossing symmetry means that the $s$-channel, $t$-channel and $u$-channel decompositions agree.
Let us write the $s$- and $t$-channel decompositions:
\begin{align}
 \left<V_1V_2V_3V_4\right>_p = \sum_{k\in \mathcal{S}^{\left<12|34\right>}} D_{k}^{\left<12|34\right>_p} \mathcal{G}_k^{\left<12|34\right>}
=
 \sum_{k\in \mathcal{S}^{\left<1|23|4\right>}} D_{k}^{\left<1|23|4\right>_p} \mathcal{G}_k^{\left<1|23|4\right>}\ .
 \label{cross}
\end{align}
In these equations:
\begin{itemize}
 \item The notation $\left<12|34\right>$ and $\left<1|23|4\right>$ mean $s$-channel and $t$-channel respectively.
 \item $D_{k}^{\left<12|34\right>_p}$ is an $s$-channel $4$-point structure constant, which does not depend on $x_1,x_2,x_3,x_4$.
 \item $\mathcal{S}^{\left<12|34\right>}\subset \mathcal{S}$ is the $s$-channel spectrum. We take it to be $p$-independent, by allowing some structure constants to vanish if necessary.
 \item $\mathcal{G}_k^{\left<12|34\right>}$ is an $s$-channel conformal block, which depends on $x_1,x_2,x_3,x_4$ but not on the extra parameter $p$.
\end{itemize}
Crossing symmetry constrains the spectrums and structure constants. We now assume that the spectrums are known, and focus on the structure constants.
Let us introduce the fusing matrix $F_{k,\ell}\left[\begin{smallmatrix} 2 & 3 \\ 1 & 4\end{smallmatrix}\right] $, which relates the $s$- and $t$-channel conformal blocks:
\begin{align}
 \mathcal{G}^{\left<12\middle| 34\right>}_k = \sum_{\ell\in \overline{\mathcal{S}}}  F_{k,\ell}\left[\begin{smallmatrix} 2 & 3 \\ 1 & 4\end{smallmatrix}\right] \mathcal{G}^{\left<1\middle|23\middle|4\right>}_\ell\ ,
\end{align}
where $\overline{\mathcal{S}}$ is a set of representations of the conformal algebra, typically much larger than the spectrum $\mathcal{S}$.
This allows us to rewrite the crossing symmetry equations \eqref{cross} as
\begin{align}
\forall \ell\in \overline{\mathcal{S}}\ , \qquad
\sum_{k\in \mathcal{S}^{\left<12\middle|34\right>}} D^{\left<12\middle|34\right>_p}_{k}  F_{k,\ell}\left[\begin{smallmatrix} 2 & 3 \\ 1 & 4\end{smallmatrix}\right] =\delta_{\ell \in \mathcal{S}^{\left<1\middle|23\middle|4\right>}} D^{\left<1\middle|23\middle|4\right>_p}_{\ell}\ .
\label{sdfd}
\end{align}
In particular,
\begin{align}
\forall \ell\in \overline{\mathcal{S}}\backslash\mathcal{S}^{\left<1\middle|23\middle|4\right>} \ , \qquad
\sum_{k\in \mathcal{S}^{\left<12\middle| 34\right>}} D^{\left<12\middle|34\right>_p}_{k}  F_{k,\ell}\left[\begin{smallmatrix} 2 & 3 \\ 1 & 4\end{smallmatrix}\right] =0\ .
\end{align}
This is a system of linear equations for the $s$-channel structure constants. The extra parameter $p$ labels linearly independent solutions $\left(D_{k}^{\left<12|34\right>_p}\right)_{k\in\mathcal{S}^{\left<12|34\right>}}$.

\subsubsection{Reduction to $4$-point correlators}

Let us show that $5$-point structure constants are determined by $4$-point structure constants.
For simplicity we assume that the crossing symmetry equations have 1d spaces of solutions, so that there are no extra parameters.
Our notations for structure constants are:
\begin{align}
 \begin{tikzpicture}[baseline = (base), scale = .25]
  \coordinate (base) at (0, -.2);
  \draw (-3, 2) node[left]{$2$} -- (-2, 0) -- node[above]{$k$} (2, 0) -- (3, 2) node[right]{$3$};
  \draw (-3, -2) node[left]{$1$} -- (-2, 0);
  \draw (2, 0) -- (3, -2) node[right]{$4$};
 \end{tikzpicture}
 \to D^{\left<12\middle|34\right>}_k
 \quad &, \quad
 \begin{tikzpicture}[baseline = (base), scale = .25]
  \coordinate (base) at (0, -.2);
  \draw (-2, -3) node[left]{$1$} -- (0, -2) -- node[left]{$k$} (0, 2) -- (-2, 3) node[left]{$2$};
 \draw (2, -3) node[right]{$4$} -- (0, -2);
  \draw (2, 3) node[right]{$3$} -- (0, 2);
 \end{tikzpicture}
 \to  D^{\left<1\middle|23\middle|4\right>}_k
 \\
 \begin{tikzpicture}[baseline = (base), scale = .32]
  \coordinate (base) at (0, -.2);
  \draw (-3, 2) node[left]{$2$} -- (-2, 0) -- node[above]{$k$} (0, 0) -- node[above]{$\ell$} (2, 0) -- (3, 2) node[right]{$4$};
  \draw (-3, -2) node[left]{$1$} -- (-2, 0);
  \draw (0, 0) -- (0, 2) node[above]{$3$};
  \draw (2, 0) -- (3, -2) node[right]{$5$};
 \end{tikzpicture}
 \to D^{\left<12\middle| 3\middle| 45\right>}_{k,\ell}
 \quad &, \quad
 \begin{tikzpicture}[baseline = (base), scale = .32]
  \coordinate (base) at (0, -.2);
  \draw (-3, 2) node[left]{$2$} -- (-2, 0) -- node[above]{$k$} (0, 0) -- node[above]{$m$} (2, 0) -- (3, 2) node[right]{$3$};
  \draw (-3, -2) node[left]{$1$} -- (-2, 0);
  \draw (0, 0) -- (0, -2) node[below]{$5$};
  \draw (2, 0) -- (3, -2) node[right]{$4$};
 \end{tikzpicture}
 \to D^{\left<12\middle| 3 4\middle|5\right>}_{k,m}
 \end{align}

\begin{prop}[Relation between $5$-point and $4$-point structure constants]
~\label{prop:54}
 If crossing symmetry equations have 1d spaces of solutions, then
 \begin{align}
  \frac{D^{\left<12|3|45\right>}_{k,\ell}}{D^{\left<k3|45\right>}_{\ell}} =\frac{D^{\left<12|3|45\right>}_{k,\ell_0}}{D^{\left<k3|45\right>}_{\ell_0}}=  \frac{D^{\left<12|34|5\right>}_{k,m}}{D^{\left<k|34|5\right>}_{m}} = \frac{D^{\left<12|34|5\right>}_{k,m_0}}{D^{\left<k|34|5\right>}_{m_0}}\ .
  \label{dddd}
\end{align}
\end{prop}
These equations state that the first ratio is $\ell$-independent, the third ratio is $m$-independent, and both ratios are equal. To prove these equations, let us apply the
correlator expansion \eqref{vovt} to a $5$-point correlator:
\begin{align}
 \left<V_1V_2V_3V_4V_5\right> = \sum_k C_k \sum_{v\in B_k} f_{k,v} \left<V_{k,v}V_3V_4V_5\right>\ .
\end{align}
We now decompose the $5$-point and $4$-point correlators into structure constants and conformal blocks, according to the expansion $V_4V_5$ or the expansion $V_3V_4$. This leads to relations for structure constants, respectively
\begin{align}
D^{\left<12|3|45\right>}_{k,\ell}= C_kD^{\left<k3|45\right>}_{\ell} \qquad , \qquad
D^{\left<12|34|5\right>}_{k,m} = C_k D^{\left<k|34|5\right>}_{m}\ .
\end{align}
Now $C_k$ may depend on all parameters of the correlators $\left<V_1V_2V_3V_4V_5\right>$ and $\left<V_{k,v}V_3V_4V_5\right>$, but neither on our choice of expansion $V_4V_5$ or $V_3V_4$, nor on
the channel representation $\ell$ or $m$. This implies Eq. \eqref{dddd}, which equates several different expressions for $C_k$.
Let us propose an alternative proof. We start from the crossing symmetry equations \eqref{sdfd}, which are the same for $5$-point structure constants as for $4$-point structure constants, and imply
\begin{align}
 \forall m\in \overline{\mathcal{S}}\backslash \mathcal{S}^{\left<k\middle|34\middle|5\right>}\ , \quad
\sum_{\ell\in \mathcal{S}^{\left<k3\middle|45\right>}} D^{\left<k3\middle|45\right>}_{\ell}  F_{\ell,m}\left[\begin{smallmatrix} 3 & 4\\ k & 5\end{smallmatrix}\right]
=\sum_{\ell\in \mathcal{S}^{\left<k3\middle|45\right>}} D^{\left<12|3\middle|45\right>}_{k,\ell}  F_{\ell,m}\left[\begin{smallmatrix} 3 & 4\\ k & 5\end{smallmatrix}\right] =0\ .
\end{align}
As functions of $\ell$, $D^{\left<k3\middle|45\right>}_{\ell}$ and $D^{\left<12|3\middle|45\right>}_{k,\ell}$ obey the same linear equations. By our assumption that these equations have 1d spaces of solutions, any two solutions must be proportional, as expressed by the first equality in \eqref{dddd}.

Of course, the $4$-point structure constants determine not only the $\ell$-dependence of $D^{\left<12|3\middle|45\right>}_{k,\ell}$, but also its $k$-dependence, as expressed by the following variant of the first equation in  Eq. \eqref{dddd}:
\begin{align}
 \frac{D^{\left<12|3|45\right>}_{k,\ell}}{D^{\left<12|3\ell\right>}_{k}} =\frac{D^{\left<12|3|45\right>}_{k_0,\ell}}{D^{\left<12|3\ell\right>}_{k_0}}\ .
\end{align}
We deduce how $D^{\left<12|3\middle|45\right>}_{k,\ell}$ depends on both $k$ and $\ell$:
\begin{align}
\frac{D_{k,\ell}^{\left<12|3\middle|45\right>}}{D_{k_0,\ell_0}^{\left<12|3\middle|45\right>}} = \frac{D_\ell^{\left<k3|45\right>}}{D_{\ell_0}^{\left<k3|45\right>}} \frac{D_k^{\left<12|3\ell_0\right>}}{D_{k_0}^{\left<12|3\ell_0\right>}} = \frac{D_\ell^{\left<k_03|45\right>}}{D_{\ell_0}^{\left<k_03|45\right>}} \frac{D_k^{\left<12|3\ell\right>}}{D_{k_0}^{\left<12|3\ell\right>}}\ .
\label{dd00}
\end{align}
This shows that our $5$-point correlator $\left<V_1V_2V_3V_4V_5\right>$ is determined by $4$-point correlators, up to a constant factor $D_{k_0,\ell_0}^{\left<12|3\middle|45\right>}$. Moreover, the second equality is a nontrivial identity for $4$-point structure constants.

\begin{prop}[Reduction to $4$-point correlators]
~\label{prop:red}
 In a conformal correlator system, all correlators are determined by $4$-point correlators, up to constant factors.
\end{prop}

This generalizes Proposition \ref{prop:54} from $5$-point correlators to $N$-point correlators. The idea is that $N$-point structure constants are combinations of $4$-point structure constants. This can be proved by performing multiple correlator expansions, or by showing that the dependence of an $N$-point structure constant on any channel field is determined by the same linear equation as the dependence of a $4$-point structure constant.

\subsubsection{Case of OPEs in CFT}

In a CFT, OPEs imply that $N$-point structure constants factorize into $2$-point structure constants $B_k$ and $3$-point structure constants $C_{ijk}$, in particular
\begin{align}
 D_k^{\left<12\middle|34\right>} = \frac{C_{12k}C_{k34}}{B_k}\quad , \quad D_{k,\ell}^{\left<12|3\middle|45\right>} = \frac{C_{12k}C_{k3\ell}C_{\ell 45}}{B_kB_\ell}\ ,
 \label{dccb}
\end{align}
which implies Eqs. \eqref{dddd} and \eqref{dd00}. In fact, the $2$-point and $3$-point structure constants can be rewritten in terms of $4$-point structure constants,
\begin{align}
 B_k = D_k^{\left<0k\middle|0k\right>} \quad , \quad C_{123} = D_3^{\left<03\middle|12\right>}\ ,
\end{align}
where $0$ stands for the identity field.

In CFTs as in conformal correlator systems, equations for structure constants become more complicated if we relax some of our technical assumptions. In particular, we could allow nontrivial field multiplicities: in the language of Definition \ref{spec} this means that the spectrum would be a multiset. Then a $4$-point structure constants would be a sum of several factorized terms, instead of being factorized as in Eq. \eqref{dccb}.

\subsubsection{Shift equations from factorizing fields}

In a conformal correlator system, there may exist some correlator expansions that are in fact OPEs. For example, in $d=2$ with local conformal symmetry, any expansion that involves the energy-momentum tensor $T$ is an OPE, and this OPE can be used for computing Virasoro blocks, including blocks that do not involve $T$ itself. In the case of critical loop models, structure constants obey shift equations that follow from OPEs of a degenerate field --- a property called interchiral symmetry \cite{rib24}. Let us generalize these examples, and show that in conformal correlator systems, the existence of OPEs constrains structure constants that may not involve these OPEs.

 \begin{defn}[Factorizing field]
 A field is factorizing if it admits OPEs with all other fields.
\end{defn}

\begin{prop}[Shift equations]
~\label{prop:shift}
Let $i$ denote a factorizing field, $1,2,3,4,j$ some other fields, and $k,\ell$ such that $V_k,V_\ell\in V_iV_j$. Then the $4$-point structure constants obey the shift equation
 \begin{align}
\rho_k=\rho_\ell \ , \quad \text{with} \quad   \rho_k = \frac{D_k^{\left<12|34\right>} D_k^{\left<ij|ij\right>}}{D_k^{\left<12|ij\right>}D_k^{\left<ij|34\right>}} \ .
\label{rhok}
 \end{align}
\end{prop}
In a CFT, we would have $\rho_k=1$, as a consequence of the factorization of $4$-point structure constants \eqref{dccb}. Let us derive the weaker property $\rho_k=\rho_\ell$, under our weaker assumptions. The idea is to introduce a field $V_p$ such that $V_2\in V_pV_i$, and to consider the following $5$-point structure constants, related by a fusion move:
\begin{align}
D_{2,k}^{\left<1|pi|34\right>}\quad & = \quad \sum_jF_{2,j}\left[\begin{smallmatrix} i & k \\ p & 1 \end{smallmatrix}\right] D^{\left<1p|i|34\right>}_{j,k} \ .
\label{d2k}
\\
 \begin{tikzpicture}[baseline = (base), scale = .35]
  \coordinate (base) at (0, -.2);
  \draw (-3, 2) node[left]{$i$} -- (-2, 0) -- node[above]{$2$} (0, 0) -- node[above]{$k$} (2, 0) -- (3, 2) node[right]{$3$};
  \draw (-3, -2) node[left]{$p$} -- (-2, 0);
  \draw (0, 0) -- (0, -2) node[below]{$1$};
  \draw (2, 0) -- (3, -2) node[right]{$4$};
 \end{tikzpicture}
 & \qquad\longrightarrow\qquad
 \begin{tikzpicture}[baseline = (base), scale = .35]
  \coordinate (base) at (0, -.2);
  \draw (-3, 2) node[left]{$p$} -- (-2, 0) -- node[above]{$j$} (0, 0) -- node[above]{$k$} (2, 0) -- (3, 2) node[right]{$3$};
  \draw (-3, -2) node[left]{$1$} -- (-2, 0);
  \draw (0, 0) -- (0, 2) node[above]{$i$};
  \draw (2, 0) -- (3, -2) node[right]{$4$};
 \end{tikzpicture}
\nonumber
\end{align}
Let us use these $5$-point structure constants for determining the $k$-dependence of $\rho_k$ \eqref{rhok}. According to Eq. \eqref{dddd},
\begin{align}
 \frac{D_k^{\left<12|34\right>}}{D_\ell^{\left<12|34\right>}}
 = \frac{D_{2,k}^{\left<1|pi|34\right>}}{D_{2,\ell}^{\left<1|pi|34\right>}}\ .
 \label{dkd2k}
\end{align}
We also use Eq. \eqref{dd00} to rewrite $D^{\left<1p|i|34\right>}_{j,k},D^{\left<1p|i|34\right>}_{j,\ell}$ in terms of the same $j$-independent quantity $D^{\left<1p|i|34\right>}_{j_0,\ell}$:
\begin{align}
 D^{\left<1p|i|34\right>}_{j,\ell} = D^{\left<1p|i|34\right>}_{j_0,\ell} \frac{D_j^{\langle 1p|i\ell\rangle}}{D_{j_0}^{\langle 1p|i\ell\rangle}}
 \quad , \quad
 D^{\left<1p|i|34\right>}_{j,k}
 = D^{\left<1p|i|34\right>}_{j_0,\ell} \frac{D_j^{\langle 1p|i\ell\rangle}}{D_{j_0}^{\langle 1p|i\ell\rangle}}  \frac{D_k^{\left<ji|34\right> }}{D_\ell^{\left<ji|34\right> }} \ .
\end{align}
Combining this with Eqs. \eqref{d2k} and \eqref{dkd2k}, we obtain
\begin{align}
 \frac{D_k^{\left<12|34\right>}}{D_\ell^{\left<12|34\right>}}
 = \frac{\sum_j F_{2,j}\left[\begin{smallmatrix} i & k \\ p & 1 \end{smallmatrix}\right]D_j^{\langle 1p|i\ell\rangle} \frac{D_k^{\left<ji|34\right> }}{D_\ell^{\left<ji|34\right> }}}{\sum_j F_{2,j}\left[\begin{smallmatrix} i & \ell \\ p & 1 \end{smallmatrix}\right] D_j^{\langle 1p|i\ell\rangle}}
 = \frac{B_\ell}{B_k} \frac{C_{k34}}{C_{\ell 34}}\frac{\sum_j F_{2,j}\left[\begin{smallmatrix} i & k \\ p & 1 \end{smallmatrix}\right]D_j^{\langle 1p|i\ell\rangle} \frac{C_{jik}}{C_{ji\ell}}}{\sum_j F_{2,j}\left[\begin{smallmatrix} i & \ell \\ p & 1 \end{smallmatrix}\right] D_j^{\langle 1p|i\ell\rangle}} \ ,
\end{align}
where the last equality uses the factorization \eqref{dccb} of $D_k^{\left<ji|34\right>}$ into $2$-point and $3$-point structure constants , which follows from the OPE $V_jV_i$. This shows that in the ratio $\frac{D_k^{\left<12|34\right>}}{D_\ell^{\left<12|34\right>}}$, the dependences on fields $V_1,V_2$ and $V_3,V_4$ factorize, which implies Eq. \eqref{rhok}. As a consequence, the ratio $\frac{D_k^{\left<12|34\right>}}{D_\ell^{\left<12|34\right>}}$ can be expressed in terms of $4$-point structure constants that involve the factorizing field $V_i$, and therefore factorize into $2$-point and $3$-point structure constants.

\section{Two-dimensional case}

In conformal field theory, two features make the case $d=2$ qualitatively different from $d>2$:
\begin{itemize}
 \item We have local conformal symmetry, described by the infinite-dimensional Virasoro algebra.
 \item It is feasible and natural to study CFT not only on the sphere, but also on arbitrary Riemann surfaces.
\end{itemize}
Let us argue that these features survive in two-dimensional conformal correlator systems. At the level of correlators,
local conformal symmetry means that we have decompositions into Virasoro blocks, and not just global conformal blocks. These decompositions, and the Virasoro blocks, can be derived with the help of the energy-momentum tensor $T$: a field whose OPE with any other field $V$ encodes the behaviour of $V$ under local conformal transformations \cite{rib24}. This derivation works in conformal correlator systems, thanks to our Proposition \ref{prop:shift}, if we take $i$ to be the energy-momentum tensor, and $\ell$ to be a Virasoro descendant of the primary $j=k$. Then the proposition states that the contribution of that descendant to a $4$-point correlator $\left<1234\right>$ is determined by the contribution of the primary.

To study conformal correlator systems on arbitrary Riemann surfaces, we need to reduce correlators at genus $g$ to correlators at genus $g-1$, by cutting a handle. This operation is analogous but not identical to a correlator expansion. We will introduce a dedicated axiom in Section \ref{sec:ch}, and show that the reduction to sphere $4$-point correlators (Proposition \ref{prop:red}) remains valid.

When it comes to the torus $0$-point correlator, the handle-cutting axiom will be weaker than in CFT, in particular it will allow fields to have arbitrary conformal spins. In Section \ref{sec:mono}, we will see that arbitrary spins lead to nontrivial monodromies for correlators on the Riemann sphere $\overline{\mathbb{C}}=\mathbb{C}\cup \{\infty\}$. To derive the correct constraints on these monodromies, we will have to remember that fields have nontrivial monodromies around $\infty$.

Finally, in Section \ref{sec:1pt}, we will focus on torus $1$-point correlators, and show how modular covariance constrains the spins. To accommodate arbitrary spins, we will have to slightly generalize the modular covariance equation, by allowing extra factors called multiplier systems.

\subsection{Cutting handles}\label{sec:ch}

Let $\mathbb{X}_{g,N}$ be the moduli space of genus $g$ Riemann surfaces with $N$ distinct punctures. An $N$-point correlator of genus $g$ is a complex-valued function $\left<V_1 \cdots V_N\right>^{(g)}$ on $\mathbb{X}_{g,N}$. (For simplicity we ignore extra parameters.)

\begin{ax}[Cutting a handle]
There exist an open cover $\mathbb{X}_{g,N}=\cup_i\Omega_i$, and holomorphic maps $\psi_{g,N|i}:\Omega_i\to \mathbb{X}_{g-1,N+2}$, such that any $N$-point correlator of genus $g\geq 1$ is a linear combination of $(N+2)$-point correlators of genus $g-1$ of the type
 \begin{align}
  \big<V_1 \cdots V_N\big>^{(g)}(X) \underset{X\in\Omega_i}{=} \sum_{R,v\in\mathcal{B}} H_R h_{R,v|i}(X) \big<V_{R,v}V_{R,v}V_1 \cdots V_N\big>^{(g-1)}\circ \psi_{g,N|i}(X)\ ,
  \label{gmo}
 \end{align}
where $H_R$ is a constant that may depend on $V_1,\cdots, V_N$,
and $h_{R,v|i}$ is a universal function on $\Omega_i$ that is determined by conformal symmetry from $\left<V_1 \cdots V_N\right>^{(g)}$ and $\psi_{g,N|i}$.
\end{ax}

This formulation of sewing Riemann surfaces is analogous to a correlator expansion \eqref{vovt}, with the structure constant $H_R$ being the analog of $C_R$.
Let us check that our axiom is obeyed in the case $(g,N)=(1,0)$ of a torus partition function in CFT, which a trace over the space of states:
\begin{align}
 \big<\big>^{(1)}(q) = \sum_{R,v\in\mathcal{B}} \left|q^{\Delta_{R,v}-\frac{c}{24}}\right|^2 = \sum_{R\in \mathcal{S}} \chi_R(q)\ .
 \label{oq}
\end{align}
Here $q=e^{2\pi i\tau}$ with $\tau$ the modulus of the torus $\frac{\mathbb{C}}{\mathbb{Z}+\tau\mathbb{Z}}$, while $c$ is the central charge of the Virasoro algebra, and $\chi_R$ is a character of the conformal algebra.
We want to relate the partition function to the sphere correlator of $2$ identical fields,
\begin{align}
 \big<V_{R,v}V_{R,v}\big>(z_1,z_2) = B_R f^{R,v}\left|z_{12}^{-2\Delta_{R,v}}\right|^2 \ ,
\end{align}
where $B_R$ is a $2$-point structure constant, and the universal factor $f^{R,v}$ is determined by conformal symmetry. We find that Eq. \eqref{gmo} is satisfied provided we set
\begin{align}
 H_R = \frac{1}{B_R} \quad , \quad \psi_{1,0}(q) = \big(q^{-\frac12}, 0\big) \quad , \quad h_{R,v}= \frac{\left|q^{-\frac{c}{24}}\right|^2}{f^{R,v}}\ .
 \label{hob}
\end{align}

\begin{defn}[2d conformal correlator system]
~\label{def:2dccs}
 Given a spectrum with a basis $\mathcal{B}$, a 2d conformal correlator system is a set of functions $\left<V_{R_1,v_1}V_{R_2,v_2}\cdots V_{R_N,v_N}\right>^{(g)}$ on $\mathbb{X}_{g,N}$ for any $g\in\mathbb{N}$ and $R_i,v_i\in \mathcal{B}$, such that:
\begin{itemize}
 \item The correlators are conformally covariant.
 \item Correlators with $N\geq 2$ fields admit expansions $V_{R_i,v_i}V_{R_j,v_j}$ on open subsets $\Omega_{i,j}\subset \mathbb{X}_{g,N}$, such that $\Omega_{i,j}\cap \Omega_{k,\ell}\neq \emptyset$.
 \item Correlators of genus $g\geq 1$ admit handle-cutting.
\end{itemize}
\end{defn}

\begin{defn}[2d CFT]
 A 2d CFT is a 2d conformal correlator system such that correlator expansions are OPEs, and the handle-cutting structure constants $H_R$ are given by the correlator-independent formula \eqref{hob}.
\end{defn}

In a CFT, each representation that belongs to the spectrum appears in the torus partition function \eqref{oq} with a positive integer coefficient. Therefore, modular invariance of the partition function strongly constrains the spectrum. In a conformal correlator system,
the structure constants $H_R$ can be correlator-dependent, and they may well vanish on a large part of the spectrum. Modular invariance of the torus $0$-point correlator can require that some representations are present in the spectrum, but not that a representation is absent. In particular, fields with non-integer spins can appear in the spectrum, although they cannot contribute to the torus $0$-point correlator.

Let us study the structure constants for torus $1$-point correlators $\left<V_1\right>^{(1)}$ and $2$-point correlators $\left<V_1V_2\right>^{(1)}$:
\begin{align}
\begin{tikzpicture}[baseline = (base), scale = .4]
  \coordinate (base) at (0, .5);
  \draw (0, 1.5) -- (0, 3.5) node[above] {$1$};
  \draw (0, 0) circle (1.5);
  \node[below] at (0, -1.5) {$\ell$};
 \end{tikzpicture}
\ \to\ D_{\ell}^{\left<1\right>^{(1)}}
 , \quad
 \begin{tikzpicture}[baseline = (base), scale = .4]
  \coordinate (base) at (0, .5);
  \draw (-2, 4) node[left]{$1$} -- (0, 3) -- (2, 4) node[right]{$2$};
  \draw (0, 1.5) -- node[right] {$k$} (0, 3);
  \draw (0, 0) circle (1.5);
  \node[below] at (0, -1.5) {$\ell$};
 \end{tikzpicture}
\ \to\ D_{k,\ell}^{\left<12\right>^{(1)}}
 , \quad
\begin{tikzpicture}[baseline = (base), scale = .4]
 \coordinate (base) at (0, .5);
 \draw (0, 0) circle (1.5);
 \draw (-1.1, 1) -- (-1.7, 3.2) node[above] {$1$};
 \draw (1.1, 1) -- (1.7, 3.2) node[above] {$2$};
 \node[below] at (0, -1.5) {$\ell$};
 \node[above] at (0, 1.5) {$m$};
\end{tikzpicture}
\ \to\ D_{\ell,m}^{\left<1|2\right>^{(1)}} .
\end{align}
Correlator expansion and handle-cutting lead to the following expressions for our $2$-point correlator:
\begin{align}
 \left<V_1V_2\right>^{(1)} = \sum_k C_k \sum_v f_{k,v} \left<V_{k,v}\right>^{(1)} = \sum_\ell H_\ell \sum_v h_{\ell,v}\left<V_{\ell,v}V_{\ell,v}V_1V_2\right> \ .
\end{align}
Decomposing into structure constants and conformal blocks, we deduce the relations
\begin{align}
 & D_{k,\ell}^{\left<12\right>^{(1)}} = C_k D_\ell^{\langle k\rangle^{(1)}}=H_\ell D_k^{\langle 12|\ell\ell\rangle} \quad , \quad
 D_{\ell,m}^{\left<1|2\right>^{(1)}} = H_\ell D_m^{\langle \ell 1|2\ell\rangle} =H_m D_\ell^{\langle 1m|m2\rangle}\ .
 \\
 & \hspace{2cm}
 \begin{tikzpicture}[baseline = (base), scale = .3]
  \coordinate (base) at (0, .5);
  \draw (0, 1.5) -- node[left] {$k$} (0, 3.5);
  \draw (0, 0) circle (1.5);
  \node[below] at (0, -1.5) {$\ell$};
 \end{tikzpicture}
 \hspace{1cm}
 \begin{tikzpicture}[baseline = (base), scale = .24]
  \coordinate (base) at (0, .5);
  \draw (-2, -3) node[left]{$\ell$} -- (0, -2) -- node[left]{$k$} (0, 2) -- (-2, 3) node[left]{$1$};
 \draw (2, -3) node[right]{$\ell$} -- (0, -2);
  \draw (2, 3) node[right]{$2$} -- (0, 2);
 \end{tikzpicture}
 \hspace{2cm}
 \begin{tikzpicture}[baseline = (base), scale = .25]
  \coordinate (base) at (0, .5);
  \draw (-3, 2) node[left]{$1$} -- (-2, 0) -- node[above]{$m$} (2, 0) -- (3, 2) node[right]{$2$};
  \draw (-3, -2) node[left]{$\ell$} -- (-2, 0);
  \draw (2, 0) -- (3, -2) node[right]{$\ell$};
 \end{tikzpicture}
 \hspace{.4cm}
 \begin{tikzpicture}[baseline = (base), scale = .25]
  \coordinate (base) at (0, .5);
  \draw (-3, 2) node[left]{$m$} -- (-2, 0) -- node[above]{$\ell$} (2, 0) -- (3, 2) node[right]{$m$};
  \draw (-3, -2) node[left]{$1$} -- (-2, 0);
  \draw (2, 0) -- (3, -2) node[right]{$2$};
 \end{tikzpicture}
 \nonumber
\end{align}
This allows us to determine the torus $2$-point structure constants $D_{k,\ell}^{\left<12\right>^{(1)}}$ in terms
of sphere $4$-point and torus $1$-point structure constants:
\begin{align}
 \frac{D_{k,\ell}^{\left<12\right>^{(1)}}}{D_{k_0,\ell_0}^{\left<12\right>^{(1)}}}
 = \frac{D_k^{\left<12|\ell\ell\right>}}{D_{k_0}^{\left<12|\ell\ell\right>}}\frac{D_\ell^{\left<k_0\right>_1}}{D_{\ell_0}^{\left<k_0\right>_1}}
 =  \frac{D_k^{\left<12|\ell_0\ell_0\right>}}{D_{k_0}^{\left<12|\ell_0\ell_0\right>}}\frac{D_\ell^{\left<k\right>^{(1)}}}{D_{\ell_0}^{\left<k\right>^{(1)}}}\ .
 \label{t2t1}
\end{align}
We also determine the structure constants $D_{\ell,m}^{\left<1|2\right>^{(1)}}$ in terms
of sphere $4$-point structure constants:
\begin{align}
 \frac{D_{\ell,m}^{\left<1|2\right>^{(1)}}}{D_{\ell_0,m_0}^{\left<1|2\right>^{(1)}}}
 = \frac{D_{\ell}^{\left<1m|m2\right>}}{D_{\ell_0}^{\left<1m|m2\right>}}
 \frac{D_{m}^{\left<\ell_01|2\ell_0\right>}}{D_{m_0}^{\left<\ell_01|2\ell_0\right>}}
 =\frac{D_{\ell}^{\left<1m_0|m_02\right>}}{D_{\ell_0}^{\left<1m_0|m_02\right>}}
 \frac{D_{m}^{\left<\ell 1|2\ell\right>}}{D_{m_0}^{\left<\ell 1|2\ell\right>}}\ .
 \label{t2s4}
\end{align}
This allows us to deduce torus $2$-point correlators from sphere $4$-point correlators. Let us also deduce torus $1$-point correlators.
Using the relation
$D_{k,\ell}^{\left<12\right>^{(1)}}=\sum_m D_{\ell, m}^{\left<1|2\right>^{(1)}} F_{m,k}\left[\begin{smallmatrix} 1 & 2 \\ \ell & \ell \end{smallmatrix}\right] $
between the two families of torus $2$-point structure constants, together with the second expression for $D_{\ell,m}^{\left<1|2\right>^{(1)}}$ in Eq. \eqref{t2s4}, and the relation
$D_{k}^{\left<\ell|12|\ell\right>}=\sum_m D_{m}^{\left<\ell 1|2\ell\right>} F_{m,k}\left[\begin{smallmatrix} 1 & 2 \\ \ell & \ell \end{smallmatrix}\right] $ from Eq. \eqref{sdfd}, we obtain
\begin{align}
 D_{k,\ell}^{\left<12\right>^{(1)}} = D_{\ell_0,m_0}^{\left<1|2\right>^{(1)}} \frac{D_{\ell}^{\left<1m_0|m_02\right>}}{D_{\ell_0}^{\left<1m_0|m_02\right>}} \frac{D_k^{\left<\ell|12|\ell\right>}}{D_{m_0}^{\left<\ell 1|2\ell\right>}}\ .
\end{align}
We use this to evaluate the right-hand side of $\frac{D_\ell^{\left<k\right>^{(1)}}}{D_{\ell_0}^{\left<k\right>^{(1)}}}= \frac{D_{k,\ell}^{\left<12\right>^{(1)}}}{D_{k,\ell_0}^{\left<12\right>^{(1)}}}$, which is
the case $k=k_0$ of Eq. \eqref{t2t1}, and we obtain
\begin{align}
\frac{D^{\left<k\right>^{(1)}}_\ell}{D^{\left<k\right>^{(1)}}_{\ell_0}}
= \frac{D_{\ell}^{\left<1m_0|m_02\right>}}{D_{\ell_0}^{\left<1m_0|m_02\right>}}
\frac{D_{m_0}^{\left<\ell_0 1|2\ell_0\right>}}
{D_{m_0}^{\left<\ell 1|2\ell\right>}}
\frac{D_k^{\left<\ell|12|\ell\right>}}{D_k^{\left<\ell_0|12|\ell_0\right>}}
\ ,
\label{t1s4}
\end{align}
which expresses torus $1$-point structure constants in terms of sphere $4$-point structure constants. The right-hand side involves 3 auxiliary fields $V_1,V_2,V_{m_0}$ that are absent from the $1$-point structure constant and that can be chosen arbitrarily. However, choosing $V_1$ to be the identity field would imply $\ell=m_0=\ell_0$, and reduce the formula to $1=1$.

Therefore, the reduction to sphere $4$-point correlators (Proposition \ref{prop:red}) applies to torus $1$-point and $2$-point correlators. By similar arguments, it also applies to any correlator in a 2d conformal correlator system.

\subsection{Monodromies of sphere $N$-point correlators}\label{sec:mono}

We have defined sphere $N$-point correlators as functions on a space $\mathbb{X}_N=\mathbb{X}_{0,N}$ \eqref{xn}.
If we allow nontrivial monodromies, correlators become multivalued functions on that space, or equivalently functions on its universal covering space $\pi:\widetilde{\mathbb{X}}_N\to \mathbb{X}_N$.

Let $\Omega\subset \mathbb{X}_N$ be a dense, open, simply connected subset of $\mathbb{X}_N$. Let $\sqcup_{i\in I} \Omega_i\subset \widetilde{\mathbb{X}}_N$ be a dense open subset of $ \widetilde{\mathbb{X}}_N$ such that $\pi\big|_{\Omega_i}$ is a biholomorphism, then $\Omega_i$ is called a sheet.
The fundamental group $\pi_1(\mathbb{X}_N)$ acts on the set of sheets, such that $g(\Omega_i) = \Omega_{g(i)}$ for $g\in \pi_1(\mathbb{X}_N)$.

For any function $f$ on $ \widetilde{\mathbb{X}}_N$ we define a family of functions $(f_i)_{i\in I}$ on $\Omega$ such that $\forall x\in \Omega_i, f_i \circ \pi(x) = f(x)$. Let $\mathcal{F}_f=\operatorname{Span}(f_i)_{i\in I}$, then $g\cdot f_i = f_{g(i)}$ defines a natural action of $\pi_1(\mathbb{X}_N)$ on $\mathcal{F}_f$. If all $f_i$ are linearly independent, $\pi_1(\mathbb{X}_N)$ acts on $\mathcal{F}_f$ by permutations, just like it acts on the sheets.
But if $\dim\mathcal{F}_f<\infty$, then the action of $\pi_1(\mathbb{X}_N)$ on $\mathcal{F}_f$ is described by finite-dimensional monodromy matrices.

\begin{defn}[Abelian monodromies]
 A correlator $f$ has abelian monodromies if $\dim\mathcal{F}_f=1$. It has finite monodromies if $\dim\mathcal{F}_f<\infty$.
\end{defn}

Let us study how global conformal symmetry constrains monodromies of sphere $N$-point correlators. We parametrize the 2d sphere using a complex coordinate $z\in \overline{\mathbb{C}}=\mathbb{C}\cup\{\infty\}$. We immediately encounter a subtlety:
conformal fields have nontrivial monodromies around $\infty$, even if there is no field at $\infty$. A primary field $V(z)$ indeed behaves as
\begin{align}
 V(z) \underset{z\to \infty}{\sim} \left|z^{-2\Delta}\right|^2 V(\infty)\ .
\end{align}
If $z$ performs an anticlockwise loop around $\infty$ i.e. $z\to ze^{-2\pi i}$, then $V(z)$ picks up a factor
\begin{align}
 M_\infty = \theta^2\ ,
\end{align}
where $\theta$ is the phase of $V(z)$. We interpret this factor as a diagonal monodromy matrix, for any correlator that includes $V(z)$.

Consider elements $g_{1,2}, \cdots , g_{1,N},g_{1,\infty}\in\pi_1(\mathbb{X}_N)$
that move $z_1$ around $z_2,\cdots ,z_N,\infty$, such that $g_{1,2}g_{1,3}\cdots g_{1,N}g_{1,\infty} = 1$:
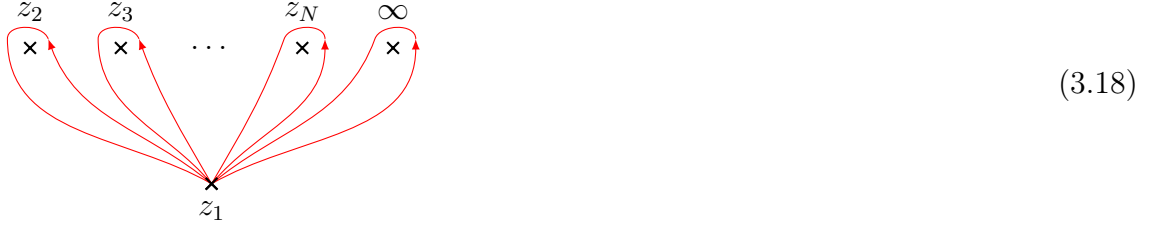
\begin{figure}[H]
 \centering
 \begin{tikzpicture}[baseline=(base), scale = .6]
 \coordinate (base) at (0, -1);
\draw (0, -3) coordinate (a) node[cross]{};
\draw (-4, 0) node[cross]{};
\draw (-2, 0) node[cross]{};
\draw (2, 0) node[cross]{};
\draw (4, 0) node[cross]{};
\node at (0, 0) {$\dots $};
\node at (-4, .8) {$z_2$};
\node at (-2, .8) {$z_3$};
\node at (2, .8) {$z_N$};
\node at (4, .8) {$\infty$};
\node at (0, -3.6) {$z_1$};
\draw[red] (-3.6, .2) to [out = 110, in = 90] (-4.5, .2) to [out = -90, in = 150] (a);
\draw[red, -latex] (a) to [out = 140, in = -70] (-3.6, .2);
\draw[red] (-1.6, .2) to [out = 110, in = 90] (-2.5, .2) to [out = -90, in = 130] (a);
\draw[red, -latex] (a) to [out = 120, in = -70] (-1.6, .2);
\draw[red] (a) to [out = 40, in = -110] (3.6, .2) to [out = 70, in = 90] (4.5, .2);
\draw[red, latex-] (4.5, .2) to [out = -90, in = 30] (a);
\draw[red] (a) to [out = 60, in = -110] (1.6, .2) to [out = 70, in = 90] (2.5, .2);
\draw[red, latex-] (2.5, .2) to [out = -90, in = 50] (a);
\draw (0, -3) node[cross]{};
 \end{tikzpicture}
 \refstepcounter{equation}
 \caption{$N$ elements of the first fundamental group, whose produce is the identity}
\end{figure}
Given an $N$-point correlator $\left<V_1(z_1)\cdots V_N(z_N)\right>$,
let $M_{1,2},M_{1,3},\cdots , M_{1,N}$ be the corresponding monodromy matrices, then
\begin{align}
 M_{1,2}M_{1,3}\cdots M_{1,N} = \theta_1^{-2}\ .
 \label{mmmt}
\end{align}
For $N=1,2,3$, conformal symmetry completely determines $N$-point correlators up to constant prefactors. Let us determine the monodromies of these correlators. In each case, we find abelian monodromies that obey the constraint \eqref{mmmt}.
\begin{enumerate}
 \item $\left< V_1(z_1)\right>\neq 0 \implies \theta_1=1$.
 \item $\left<V_1(z_1)V_2(z_2)\right> \propto \left|z_{12}^{-2\Delta_1}\right|^2$ with $\theta_1=\theta_2$, therefore $M_{1,2}= \theta_1^{-2}$.
 \item $
  \left<V_1(z_1)V_2(z_2)V_3(z_3)\right>\propto \left|z_{12}^{\Delta_3-\Delta_1-\Delta_2}z_{13}^{\Delta_2-\Delta_1-\Delta_3} z_{23}^{\Delta_1-\Delta_2-\Delta_3}\right|^2
 $ has the monodromies
 \begin{align}
  M_{1,2} = \theta_3\theta_1^{-1}\theta_2^{-1} \quad , \quad M_{1,3} = \theta_2\theta_1^{-1}\theta_3^{-1} \quad , \quad M_{2,3} = \theta_1\theta_2^{-1}\theta_3^{-1}\ .
 \end{align}
 (Single-valuedness $M_{i,j}=1$ would imply $\theta_i\in \{-1,1\}$ with $\theta_1\theta_2\theta_3=1$.)
\end{enumerate}
An $N\geq 4$-point correlator is not fully determined by conformal symmetry, and we have nontrivial monodromy constraints on spectrums and structure constants.

In the case of abelian monodromies, let us call
 $\theta_{i,j}=\theta_{j,i}$ the channel phase that appears in the expansion $V_iV_j$ with $i\neq j$.
 (In Eq. \eqref{tk12}, this would amount to renaming $\theta_k=\theta_{1,2}$.)
 These $\frac12 N(N-1)$ channel phases are subject to $N$
 monodromy constraints of the type \eqref{mmmt}, where the monodromy matrices are diagonal $M_{i,j}=\theta_{i,j}\theta_i^{-1}\theta_j^{-1}$. The constraints therefore amount to
 \begin{align}
\forall i=1,2,\cdots, N\  , \quad  \prod_{1\leq j\neq i\leq N} \theta_{i,j} =  \theta_i^{N-4} \prod_{i=1}^N\theta_i\ .
\end{align}
In the case $N=4$, this implies $\theta_{1,2}^2=\theta_{3,4}^2$. From the expansions $V_1V_2$ and $V_3V_4$ and conformal symmetry, we deduce the slightly stronger condition $\theta_{1,2}=\theta_{3,4}$, allowing us to define the channel phases
\begin{align}
 \theta_s = \theta_{1,2}=\theta_{3,4} \quad , \quad \theta_t=\theta_{1,4}=\theta_{2,3} \quad , \quad \theta_u=\theta_{1,3}=\theta_{2,4}\ .
\end{align}
We have used $3$ of our $4$ constraints to define these channel phases, and the remaining constraint amounts to Eq. \eqref{tstttu}, whose full statement is:

\begin{prop}[Sphere $4$-point correlators with abelian monodromies]
A sphere $4$-point correlator has abelian monodromies if and only if all fields in any given channel $x\in\{s,t,u\}$ have the same phase $\theta_x$. In terms of phases or spins, the monodromy constraint is
 \begin{align}
  \theta_s\theta_t\theta_u = \theta_1\theta_2\theta_3\theta_4\ \iff \
  S_s+S_t+S_u = S_1+S_2+S_3+S_4   \bmod \mathbb{Z}\ .
  \label{sisx}
 \end{align}
\end{prop}
\vspace{2mm}

Since correlators now live on the universal covering space $\widetilde{\mathbb{X}}_N$ instead of $\mathbb{X}_N$ \eqref{xn}, some of our axioms and definitions have to be modified. While $\mathbb{X}_N$ has a natural action of the permutation group, $\widetilde{\mathbb{X}}_N$ has an action of the braid group, and we can no longer assume permutation invariance as in the original definition \ref{def:cor} of correlators. Moreover the notion of coinciding points that appears in Axiom \ref{ax:ce} becomes more subtle. We will not try to give an appropriate reformulation of the axiom, see however Section \ref{sec:fb} for the example of free bosonic correlators.

\subsection{Modular covariance of torus $1$-point correlators}\label{sec:1pt}

Let $\frac{\mathbb{C}}{\mathbb{Z}+\tau\mathbb{Z}}$ be a torus of modulus $\tau$. An element $g=\left(\begin{smallmatrix} a& b\\ c & d\end{smallmatrix}\right)\in SL_2(\mathbb{Z})$ of the modular group acts on $\tau$ as $g\cdot \tau = \frac{a\tau+b}{c\tau+d}$. The modular group is generated by $S=\left(\begin{smallmatrix} 0& -1\\ 1 & 0\end{smallmatrix}\right)$ and $T=\left(\begin{smallmatrix} 1& 1\\ 0 & 1\end{smallmatrix}\right)$, where $S^2 = \left(\begin{smallmatrix} -1& 0\\ 0 & -1\end{smallmatrix}\right)$ acts trivially $S^2\cdot \tau = \tau$.

\begin{defn}[Modular covariance]
 The function $Z(\tau)$ is modular covariant with weights $\Delta_1,\bar\Delta_1$ if
 \begin{align}
 Z(g\cdot \tau) = \epsilon(g) \left|(c\tau +d)^{\Delta_1}\right|^2 Z(\tau)\ ,
\end{align}
for some function $\epsilon$ on $SL_2(\mathbb{Z})$ called a multiplier system. Then the phase $\theta_1=e^{2\pi i (\Delta_1-\bar\Delta_1)}$ obeys
\begin{align}
 \epsilon(S^2)\sqrt{\theta_1}=1\ .
 \label{es2}
\end{align}
\end{defn}

If the multiplier system was trivial $\epsilon =1$, this would reduce to $\sqrt{\theta_1}=1$, and the spin $\Delta_1-\bar\Delta_1$ would have to be an even integer. So we need multiplier systems to deal with arbitrary spins. A precise definition of modular covariance and multiplier systems can be found in \cite{pas03}. In particular, the action $Z((g_1g_2)\cdot \tau) = Z(g_1\cdot (g_2\cdot \tau))$ of $SL_2(\mathbb{Z})$ does not necessarily imply $\epsilon(g_1g_2)=\epsilon(g_1)\epsilon(g_2)$, due to the non-integer power in $(c\tau +d)^{\Delta_1}$.

\begin{ax}[Modular covariance of torus $1$-point correlators]
 A torus $1$-point correlator $\left<V_1\right>^{(1)}$ of a primary field is modular covariant with weights $\Delta_1,\bar\Delta_1$.
\end{ax}

This axiom can be viewed as part of the definition \ref{def:2dccs} of 2d conformal correlator systems, where it contributes to the technical statement of conformal covariance. Until now we did not need the precise meaning of conformal covariance, as it was the same in conformal correlator systems as in CFT. This is no longer the case with modular covariance. It would be nice to deduce our axiom from the more fundamental principle of Weyl covariance \cite{gkrv21}.

Now, we allow correlators to have nontrivial monodromies as functions of $\tau$, so our torus $1$-point correlator is a function on the universal covering space $\widetilde{\mathbb{X}}_{1,1}$.
Let us decompose a torus $1$-point correlator into conformal blocks:
\begin{align}
 \left<V_1\right>^{(1)} = \sum_k D_k^{\left<1\right>^{(1)}} \left|\frac{q^{\Delta_k-\frac{c-1}{24}}}{\eta(q)} \left(1+\sum_{j=1}^\infty a_j q^j\right)\right|^2\ ,
\end{align}
where $a_j$ is a known function of $c,\Delta_1,\Delta_k$, we define $q=e^{2\pi i\tau}$, and we use the Dedekind eta function
\begin{align}
 \eta(q)=q^{\frac{1}{24}}\prod_{j=1}^\infty (1-q^j)\ .
 \label{ded}
\end{align}
The monodromy of our $1$-point correlator around $q=0$ corresponds to the modular transformation $T\cdot q = qe^{2\pi i}$, and
is determined by the channel phases $\theta_k$. Let us focus on the case of abelian monodromy, when the channel phase $\theta_k=\theta$ is $k$-independent. By modular covariance, we therefore have
\begin{align}
 \epsilon(T) = \theta\ .
 \label{et}
\end{align}
The structure of the modular group, in particular $(ST)^3=1$, implies relations between $\epsilon(S),\epsilon(T)$ and $\epsilon(S^2)$ \cite{pas03}
\begin{align}
 \epsilon(S)=\epsilon(T)^{-3}\qquad , \qquad \epsilon(S^2)=\epsilon(S)^2\ .
\end{align}
For example, the Dedekind eta function \eqref{ded} is modular covariant with $\epsilon(T)=e^{i\frac{\pi}{12}}$ and $\epsilon(S)=e^{-i\frac{\pi}{4}}$, it has weights $(\Delta_1,\bar\Delta_1)=(\frac12, 0)$, and abelian monodromies. In general, we obtain the relation \eqref{tt1} between the phases $\theta$ \eqref{et} and $\theta_1$ \eqref{es2}, whose full statement is:

\begin{prop}[Torus $1$-point correlators with abelian monodromies]
A torus $1$-point correlator has abelian monodromies if and only if all channel fields have the same phase $\theta$. In terms of phases or spins, the monodromy constraint is
 \begin{align}
  \theta_1= \theta^{12}
  \ \iff \
  S_1=12S   \bmod \mathbb{Z}\ .
  \label{tot}
 \end{align}
\end{prop}
This can actually be derived from the case of sphere $4$-point correlators \eqref{sisx}, using the sphere-torus relation \cite{rrj26}. A torus $1$-point correlator indeed corresponds to a sphere $4$-point correlator with phases $(\theta'_1,\theta'_2,\theta'_3,\theta'_4)=(\sqrt{\theta_1},1,1,1)$ and $(\theta'_s,\theta'_t,\theta'_u)=(\theta^2,\theta^2,\theta^2)$.

As a consequence, in a conformal correlator system with nontrivial abelian monodromies, the phase must take infinitely many values. If we indeed have a field $V_1$ with $\theta_1\neq 1$, then fields with phase $\theta_1^\frac{1}{12}$ must appear as channel fields in $\left<V_1\right>^{(1)}$. Considering torus $1$-point correlators of fields with phase $\theta_1^\frac{1}{12}$ leads to channel fields with phase $\theta_1^{\frac{1}{144}}$, etc.

For a torus $1$-point correlator with nonabelian monodromies, the action of the modular group exchanges the sheets of $\widetilde{\mathbb{X}}_{1,1}$, and the multiplier system $\epsilon$ is matrix-valued --- the counterpart of the monodromy matrices of sphere correlators.

\section{Examples} \label{sec:ex}

\subsection{Critical loop models}\label{sec:clm}

Let us review correlators in critical loop models, and argue that they do not admit OPEs, as this is our original motivation for introducing conformal correlator systems. For more details see \cite{rib24}.

We write $V_{(r,s)}$ for a primary field of left and right dimensions $(\Delta,\bar\Delta)=(\Delta_{(r,s)},\Delta_{(r,-s)})$, where $r,s$ are Kac indices, which means
\begin{align}
\Delta_{(r,s)} = \frac14\left(\beta r-\beta^{-1}s\right)^2 -\frac14 \left(\beta -\beta^{-1}\right)^2
\qquad \text{with} \qquad
 c= 13-6\beta^2-6\beta^{-2} \ .
 \label{cdp}
\end{align}
The spectrum of primary fields is determined by two constraints:
\begin{multicols}{2}
\begin{itemize}
 \item $S=rs\in \mathbb{Z}$, the spin is integer.
 \item $2r\in\mathbb{N}$ is the number of legs.
\end{itemize}

 $
 \begin{tikzpicture}[baseline=(base), scale = .8]
 \coordinate (base) at (0, 0);
\draw[thick, red] plot [smooth] coordinates {(0, 0)(.2, .7)(.4, 1.2)};
\node at (.6, 1.3) {$\scriptstyle{1}$};
\draw[thick, red] plot [smooth] coordinates {(0, 0)(.3, .6)(.6, 1)};
\node at (.8, 1.1) {$\scriptstyle{2}$};
\draw[thick, red] plot [smooth] coordinates {(0, 0)(.4, .5)(.8, .8)};
\node at (1, .9) {$\scriptstyle{3}$};
\draw[thick, red] plot [smooth] coordinates {(0, 0)(.5, -.3)(1.1, -.5)};
\node at (1.4, -.5) {$\scriptstyle{2r}$};
 \draw (0, 0) node[fill, circle, minimum size = 1.6mm, inner sep = 0]{};
 \node[left] at (0, 0) {$V_{(r,s)}$};
 \draw[dashed] plot [smooth] coordinates {(.6, .45)(.75, .1)(.7, -.25)};
\end{tikzpicture}
$
\end{multicols}
\noindent
Critical loop models include the critical $O(n)$, $Q$-state Potts and $PSU(n)$ models, with parameters $n=-2\cos(\pi\beta^2)$ and $Q=4\cos^2(\pi\beta^2)$.

\subsubsection{Necessity of complex spins}

In critical loop models, a number of structure constants are exactly known, and they are written in terms of the double Gamma function. In particular, up to sign factors, the 3-point structure constants are \cite{nrj23}
\begin{align}
 C_{123} \propto \prod_{\epsilon_i=\pm} \Gamma_\beta^{-1}\left(\tfrac{\beta+\beta^{-1}}{2} + \tfrac{\beta}{2}\left|\textstyle{\sum}_{i=1}^3 \epsilon_i r_i\right| + \tfrac{\beta^{-1}}{2} \textstyle{\sum}_{i=1}^3\epsilon_is_i\right)\ .
 \label{cott}
\end{align}
This formula was guessed from special cases, by taking inspiration from Liouville theory, whose structure constants are given by the same formula with $r_i=0$. At the time of this writing, no derivation is known.

To solve Liouville theory, we can use crossing symmetry of 4-point functions that involve the degenerate fields $V^d_{\langle 1,2\rangle}$ and $V^d_{\langle 2,1\rangle}$. These degenerate fields determine how structure constants behave under $s_i\to s_i+2$ and $s_i\to s_i+2\beta^2$ respectively. The resulting shift equations can be solved in terms of the double Gamma function $\Gamma_\beta(x)$, which is characterized by its behaviour under $x\to x+\beta$ and $x+\beta^{-1}$.

In critical loop models, the degenerate field $V^d_{\langle 1,2\rangle}$ is known to exist. The same analytic bootstrap approach as in Liouville theory constrains the shifts $s_i\to s_i+2$, which are compatible with the integer spin condition $r_is_i\in\mathbb{Z}$. However, to uniquely determine structure constants, we would also need the field $V^d_{\langle 2,1\rangle}$, which would lead to shifts of the type $r_i\to r_i+2$ or $s_i\to s_i+2\beta^2$. Given a primary field $V_{(r,s)}$ with integer spin $rs\in\mathbb{Z}$, such shifts would make the spin rational or complex respectively.

Therefore, complex spins are necessary for determining 3-point structure constants by solving shift equations. They would also be necessary for determining them from their zeros, since $\Gamma_\beta^{-1}(x)$ has zeros for $x\in -\beta\mathbb{N}-\beta^{-1}\mathbb{N}$. The 3-point structure constants do not have any remarkable property if spins are integer (or rational, or real): complex spins are more natural.

\subsubsection{No OPEs in critical loop models}

A sphere $N$-point correlator depends not only on $N$ primary fields, but also on a combinatorial map $M$, which describes how their legs are connected. In our language, $M$ is an extra parameter. In the case $N=3$, the combinatorial map is unique, and we know the $3$-point structure constant \eqref{cott}. In the case $N=4$, the $s$-channel decomposition is of the type
\begin{align}
 \left<V_1V_2V_3V_4\right>_M = \sum_k D_k^{\left<12|34\right>_M} \mathcal{G}_k^{\left<12|34\right>}\ .
\end{align}
The $4$-point structure constants $D_k^{\left<12|34\right>_M}$ are known analytically in quite a few examples: in general they depend non-trivially on $M$, and do not factorize into $3$-point structure constants as in Eq. \eqref{dccb}:
\begin{align}
 D_k^{\left<12|34\right>_M} \neq \frac{C_{12k}C_{k34}}{B_k} \ .
\end{align}
The dependence on $M$, and the failure of factorization, are not enough to conclude that OPEs cannot exist.
We would also expect such features in the presence of OPEs, if we had nontrivial field multiplicities: then a $4$-point structure constant would be a sum over the $s$-channel fields $V_k^\nu$ that transform in a given representation $R_k$ of the conformal algebra, of the type $D_k = \sum_\nu \frac{C_{12(k,\nu)}C_{(k,\nu)34}}{B_{(k,\nu)}}$.

In the critical $O(n)$ model with $n\in\mathbb{N}_{\geq 2}$, we could define $\nu$ as labelling a component of an $O(n)$ tensor. For example, for the primary field $V_{(\frac12,0)}$, we would have $\nu=1,2,\cdots, n$. All correlators would then be dressed with $O(n)$ tensors, and we would expect that they admit OPEs. However, this description of correlators would be excessively complicated, since there are many more $O(n)$ tensors than combinatorial maps \cite{gjnrs23}. And no such description is known for $n\notin\mathbb{N}_{\geq 2}$.

As this example shows, we need to fix the spectrum before we can disprove the existence of OPEs. It may always be possible to enlarge the spectrum by adding field multiplicities, such that all correlator expansions are OPEs. But this would not necessarily lead to a simpler description of the correlator system.

\subsubsection{Deformed critical loop models}

For a given phase $\theta\in\mathbb{C}^*$, let us consider the set of primary fields
\begin{align}
 \mathcal{B}_\theta = \left\{V_{(r,s)}\right\}_{\substack{r\in\frac12\mathbb{N}^* \\ e^{2\pi i rs}=\theta}}\ .
\end{align}
For $\theta=1$, this coincides with the set of primary fields in critical loop models, minus the fields $V_{(0,s)}$ (called diagonal fields):
\begin{align}
 \mathcal{B}_1 = \left\{V_{(r,s)}\right\}_{\substack{r\in\frac12\mathbb{N}^* \\ rs\in\mathbb{Z}}}\ .
 \label{b1}
\end{align}
For $\theta\neq 1$, the set $\mathcal{B}_\theta$ is a non-integer-spin deformation of $\mathcal{B}_1$. Since all fields have the same phase, $\mathcal{B}_\theta$ leads to correlators with abelian monodromies.

\begin{conj}[Sphere $4$-point correlators with abelian monodromies]
Consider a sphere $4$-point correlator of primary fields $\left<V_1V_2V_3V_4\right>$, and phases $\theta_s,\theta_t,\theta_u$ that obey the constraint \eqref{tstttu}. The space of solutions of crossing symmetry equations
\begin{align}
 \mathcal{Z}_{\left<V_1V_2V_3V_4\right>} = \left\{\left(D_k^{(x)}\right)_{\substack{x\in\{s,t,u\}\\ k\in \mathcal{B}_{\theta_x}}}
 \middle| \sum_{k\in\mathcal{B}_{\theta_s}} D_k^{(s)}\mathcal{G}_k^{(s)}
 = \sum_{k\in\mathcal{B}_{\theta_t}} D_k^{(t)}\mathcal{G}_k^{(t)}
 = \sum_{k\in\mathcal{B}_{\theta_u}} D_k^{(u)}\mathcal{G}_k^{(u)}
 \right\}
\end{align}
has the finite dimension
\begin{align}
 \dim \mathcal{Z}_{\left<V_1V_2V_3V_4\right>} = \left\lfloor r_1^2+r_2^2+r_3^2+r_4^2-\frac12\right\rfloor\ .
\end{align}
\end{conj}
If $\theta_i=1$ for all $i\in\{1,2,3,4,s,t,u\}$, this reduces to a well-tested conjecture in critical loop models. In particular, $\dim\mathcal{Z}_{\left<V_1V_2V_3V_4\right>}$ is the number of weakly connected combinatorial maps with vertices of valencies $2r_1,2r_2,2r_3,2r_4$ \cite[Eq. (2.13)]{gjnrs23}. We now conjecture that this dimension does not change if we allow nontrivial phases, provided they obey the constraint \eqref{tstttu}. In other words, the correlators become functions of $7$ phases, subject to $1$ constraint.

The validity of this conjecture is supported by numerically solving crossing symmetry equations for various $4$-point correlators. The solutions that we find are examples of correlators with nontrivial abelian monodromies, which can be computed numerically to any given precision.
Moreover, if the constraint \eqref{tstttu} is violated, we do not find any solution.

\begin{conj}[Torus $1$-point correlators with abelian monodromies]
Consider a torus $1$-point correlator of a primary field $\left<V_1\right>^{(1)}$ with $r_1\in\mathbb{N}^*$, and a phase $\theta\neq 1$ that obeys the constraint \eqref{tt1}. The space of solutions of modular covariance equations
\begin{align}
 \mathcal{Z}_{\left<V_1\right>^{(1)}} = \left\{\left(D_k\right)_{k\in \mathcal{B}_{\theta}}
 \middle| \forall g_1, g_2\in PSL_2(\mathbb{Z}) , \ \ \sum_{k\in\mathcal{B}_{\theta}} D_k\mathcal{G}_k^{(g_1)}
 = \sum_{k\in\mathcal{B}_{\theta}} D_k\mathcal{G}_k^{(g_2)}
 \right\}
\end{align}
has the finite dimension
\begin{align}
 \dim \mathcal{Z}_{\left<V_1\right>^{(1)}} = \left\lfloor \frac{r_1^2+9}{6}\right\rfloor+\delta_{r_1\equiv 0\bmod 6}-2\ .
 \label{dz11}
\end{align}
\end{conj}
In the limit $\theta=\theta_1=1$, this does not quite reduce to the well-tested conjecture of \cite{rrj26}. According to \cite[Eq. (2.11)]{rrj26}, there are indeed $\left\lfloor \frac{r_1^2+9}{6}\right\rfloor+\delta_{r_1\equiv 0\bmod 6}$ relevant combinatorial maps. We lose one solution because we exclude diagonal fields from the channel spectrum: this is analogous to restricting to weakly connected maps in the case of sphere $4$-point correlators. However, the $-2$ in Eq. \eqref{dz11} means that we lose a second solution for unknown reasons. In particular, there is no nonzero solution for $r_1=2$, and only $1$ solution for $r_1=3$.

Nevertheless, our numerical checks of the conjecture provide strong evidence that the condition \eqref{tt1} is correct, and provide examples of $1$-point torus correlators with nontrivial abelian monodromies.

Our numerical checks of the conjectures can be found in an ancillary Julia notebook, which relies on publicly available bootstrap code by Paul Roux \cite{roux252}.

\subsubsection{Extended critical loop models}

In order to build correlators with nonabelian monodromies, we need spectrums where the phase is not constant. A simple possibility is
\begin{align}
 \mathcal{B}^P = \left\{V_{(r,s)}\right\}_{\substack{r\in\frac12\mathbb{N}^* \\ rs\in\frac{1}{P}\mathbb{Z}}}\ ,
\end{align}
where $P\in\mathbb{N}^*$. Then phases are $P$-th roots of unity. The case $P=1$ reduces to the spectrum of critical loop models, $\mathcal{B}^1=\mathcal{B}_1$ \eqref{b1}. Since $\mathcal{B}^1\subset \mathcal{B}^P$, replacing $\mathcal{B}^1$ with $\mathcal{B}^P$ can only enlarge the space of solutions of conformal bootstrap equations, but does it become strictly larger?

In the case of torus $1$-point correlators, it was observed that taking $\mathcal{B}^2$ instead of $\mathcal{B}^1$ as the set of channel fields does lead to extra solutions of modular covariance equations. The extra solutions are fermionic, since the phases are in $\{-1,1\}$ \cite[Section 5.4]{rrj26}. We expect that higher values of $P$ lead to more correlators with nonabelian monodromies, which belong to extensions of critical loop models.

\subsection{Free bosons}\label{sec:fb}

In a free bosonic theory, the chiral symmetry algebra is an abelian affine Lie algebra. For any complex value of a parameter called the background charge $Q\in\mathbb{C}$, that algebra has a Virasoro subalgebra with central charge $c=1+6Q^2$. An affine primary field with momentum $\alpha\in\mathbb{C}$ is also a Virasoro primary field with conformal dimension $\Delta=\alpha(Q-\alpha)$.

The abelian affine symmetry determines correlators of affine primary fields, which are given by the simple formula
\begin{align}
 \big\langle V_1(z_1)\cdots V_N(z_N)\big\rangle = \left|\delta\big(\textstyle{\sum}_{i=1}^N\alpha_i-Q\big)\prod_{i<j} z_{ij}^{-2\alpha_i\alpha_j}\right|^2\ ,
 \label{zaa}
\end{align}
where the delta function prefactor imposes momentum conservation. In theories such as compactified free bosons, the values of the left and right momentums $\alpha_i,\bar\alpha_i$ are further constrained by single-valuedness of correlators \cite{rib14}.

Here, we do not impose single-valuedness, and allow arbitrary $\alpha_i,\bar\alpha_i\in\mathbb{C}$, subject only to momentum conservation. This gives rise to a conformal correlator system. Since any correlator expansion gives rise to only one primary field, monodromies are abelian. Therefore, any $4$-point correlator must obey the constraint \eqref{tstttu}. In fact, this follows from a stronger constraint that is obeyed independently by the left- and right-moving momentums,
\begin{align}
 \Delta_s+\Delta_t+\Delta_u=\Delta_1+\Delta_2+\Delta_3+\Delta_4\ .
\end{align}
This is a consequence $\alpha_s=\alpha_1+\alpha_2$, $\alpha_t=\alpha_1+\alpha_4$ and $\alpha_u=\alpha_1+\alpha_3$, as dictated by momentum conservation.

In order to perform correlator expansions, we have to be careful with the geometry of the space $\widetilde{\mathbb{X}}_N$ where our correlator lives, as we warned at the end of Section \ref{sec:mono}. As written in Eq. \eqref{zaa}, our correlator is unambiguous for $z_1>z_2>\cdots > z_N \in \mathbb{R}$. Taking the limit $z_i\to z_{i+1}$, we obtain an expansion with the trivial coefficient $C_k=1$, which is therefore an OPE. However, the expansion $V_1(z_1)V_3(z_3)$, computed by moving $z_1$ past $z_2$ (above or below), is not an OPE, because its coefficient $e^{\pm 2i\pi(\alpha_1\alpha_2-\bar\alpha_1\bar\alpha_2)}$ depends on the field $V_2$. This coefficient describes the action of a braid: surely the definition of OPEs should be modified, and take braiding into account. In any case, the OPEs $V_i(z_i)V_{i+1}(z_{i+1})$ are enough for reducing $N$-point correlators to $3$-point correlators. So we consider free bosons as conformal correlator systems with OPEs.

\subsection{Coulomb gas integrals}

Coulomb gas integrals are generalizations of free bosonic correlators, where the conservation of momentum is replaced by the weaker constraint
\begin{align}
 \sum_{i=1}^N \alpha_i = Q - m b \quad \text{with} \quad m\in\mathbb{N}\ ,
\end{align}
where $m=0$ corresponds to the free bosonic case, and the parameter $b$ is defined by $Q=b+b^{-1}$. We then define the correlators of diagonal affine primary fields
\begin{multline}
 \big\langle V_{\alpha_1}(z_1)\cdots V_{\alpha_N}(z_N)\big\rangle \ \underset{\sum_{i=1}^N \alpha_i = Q - m b}{=}\ \prod_{i<j} |z_{ij}|^{-4\alpha_i\alpha_j}
 \\
 \times \int_\Gamma\left(\prod_{k=1}^m  d^2y_k\right) \prod_{k=1}^m\prod_{i=1}^N |y_k-z_i|^{-4b\alpha_i} \prod_{k<\ell} |y_{k\ell}|^{-4b^2}\ ,
\end{multline}
where $\Gamma\subset \mathbb{C}^{2m}$ is a contour of complex dimension $m$ in the complexified space of values of $y_k$, such that the integral converges. For simplicity we consider integrals that only involve the screening charge $\int V_b$, and not the dual charge $\int V_{b^{-1}}$.

Coulomb gas integrals admit OPEs, according to the fusion rule
\begin{align}
 V_{\alpha_1}V_{\alpha_2} = \sum_{m=0}^\infty V_{\alpha_1+\alpha_2+mb}\ ,
 \label{smoi}
\end{align}
where only finitely many terms contribute to any given correlator. Coulomb gas integrals are single-valued if $\Gamma=\mathbb{C}^m$. There are CFTs like Liouville theory or minimal models, where some or all correlators are Coulomb gas integrals.

We can generalize Coulomb gas integrals by allowing the left- and right-moving momentums to differ, $\alpha\neq \bar\alpha$:
\begin{multline}
 \big\langle V_{\alpha_1,\bar\alpha_1}(z_1)\cdots V_{\alpha_N,\bar\alpha_N}(z_N)\big\rangle \ \underset{\sum_{i=1}^N \alpha_i = \sum_{i=1}^N \bar\alpha_i = Q - m b}{=}\ \prod_{i<j} \left|z_{ij}^{-2\alpha_i\alpha_j}\right|^2
 \\
 \times \int_\Gamma\left(\prod_{k=1}^m  d^2y_k\right) \prod_{k=1}^m\prod_{i=1}^N \left|(y_k-z_i)^{-2b\alpha_i}\right|^2 \prod_{k<\ell} |y_{k\ell}|^{-4b^2}\ ,
\end{multline}
Let us first consider the case when the integrand remains single-valued. In the case of the factor $\left|(y-z)^{-2b\alpha}\right|^2$, this implies
\begin{align}
 \alpha-\bar\alpha \in \frac12 b^{-1}\mathbb{Z}\ .
\end{align}
This condition also ensures that monodromies are abelian. And indeed, the conformal spin $S=(\alpha-\bar\alpha)(Q-\alpha-\bar\alpha)$ only changes by an integer under the shift $(\alpha,\bar\alpha)\to (\alpha+b,\bar\alpha+b)$ that corresponds to the fusion rule \eqref{smoi}.

More general values of $\alpha,\bar\alpha$ lead to non-abelian monodromies.
For example, correlators in critical XXZ$_q$ spin chains are Coulomb gas integrals \cite{ggqzz24}. These models exist for generic values of the central charge, and their primary fields are degenerate. The left-moving and right-moving Kac indices of a primary field are in general different, so its spin needs not be integer, nor even rational. OPEs exist, they are constrained by fusion rules of degenerate fields, they are known explicitly \cite[(2.16)]{ggqzz24}, and they involve sums over fields with different phases, so monodromies are non-abelian.

\subsection{Topological defects}

Let us elaborate on the example of a sphere $4$-point correlator from Figure \ref{defect}. If there was no defect, we would have a single-valued correlator. In the notations of Eq. \eqref{cross}, its $s$- and $t$-channel decompositions would read
\begin{align}
 \left<V_1V_2V_3V_4\right> = \sum_{k\in \mathcal{S}^{\left<12|34\right>}} \frac{C_{12k}C_{k34}}{B_k} \mathcal{G}_k^{\left<12|34\right>}
=
 \sum_{k\in \mathcal{S}^{\left<1|23|4\right>}} \frac{C_{14k}C_{k23}}{B_k} \mathcal{G}_k^{\left<1|23|4\right>}\ .
 \label{4pt}
\end{align}
Let us introduce a topological defect line $\mathcal{L}$ around the fields $V_1,V_2$. This does not obstruct the OPE $V_1V_2$, but this modifies the coefficients of the $s$-channel decomposition, by inserting eigenvalues of the defect operator $\widehat{\mathcal{L}} V_k =\ell_k V_k$ \cite{clswy18}:
\begin{align}
 \left<V_1V_2V_3V_4\right>_\mathcal{L} = \sum_{k\in \mathcal{S}^{\left<12|34\right>}} \frac{C_{12k}C_{k34}}{B_k}\ell_k \mathcal{G}_k^{\left<12|34\right>}\ .
\end{align}
The correlator remains invariant under the monodromy tranformation $z_{12}\to e^{2\pi i}z_{12}$, since each $s$-channel block is invariant.
On the other hand, the defect line prevents us from performing the OPE $V_1V_4$. As a result, the $t$-channel decomposition changes a lot, and we obtain an example of the general formula \eqref{cross},
\begin{align}
 \left<V_1V_2V_3V_4\right>_\mathcal{L} = \sum_{k\in \mathcal{S}^{\left<1|23|4\right>_\mathcal{L}}} D_{k}^{\left<1|23|4\right>_\mathcal{L}} \mathcal{G}_k^{\left<1|23|4\right>}\ ,
\end{align}
where the defect plays the role of an extra parameter. Both $\mathcal{S}^{\left<1|23|4\right>_\mathcal{L}}$ and $D_{k}^{\left<1|23|4\right>_\mathcal{L}}$ can be computed from defect parameters. The defect spectrum $\mathcal{S}^{\left<1|23|4\right>_\mathcal{L}}$ typically includes fields with non-integer spins.

For example, in a diagonal minimal model, the spectrum is made of diagonal fields, which we schematically write $V_{k,k}$ where $k$ labels a representation in the Kac table $K$. Introducing topological defects leads to spectrums that include all fields of the type $V_{k,\overline{k}}$, where the left-moving and right-moving representations are no longer the same. In the presence of a configuration $\mathcal{L}$ of topological defects (which may include defect junctions), sphere $4$-point correlators are therefore of the form $\sum_{k,\overline{k}\in K}D_{k,\overline{k}}^{\left<12|34\right>_\mathcal{L}} \mathcal{G}_{k,\overline{k}}^{\left<12|34\right>}$. It is not clear to us whether we this gives rise to all possible linear combinations of the conformal blocks $\mathcal{G}_{k,\overline{k}}^{\left<12|34\right>}$.
Anyway, we could forget about defects, and define a maximal conformal correlator system, by considering all solutions of BPZ equations for $N$-point correlators of fields in the Kac table, with no constraints on how left-moving and right-moving solutions are coupled.

\section{Concluding remarks}

We have introduced \textit{conformal correlator systems} in order to describe correlators that have conformal symmetry, but are not related by operator product expansions. This was necessary because \textit{conformal field theory} cannot describe such objects without losing too much of its meaning. In the example of topological defects, we could accommodate the lack of OPEs in the framework of CFT. However, in more general examples such as critical loop models, there is no reason for correlators to have an interpretation in terms of non-local objects in some CFT. Even if possible, constructing such non-local objects may well be more complicated than directly studying the correlators' properties.

\subsubsection{Terminology: CFT or not CFT?}

Let us list a number of axioms that may or may not be included in the definition of conformal field theory, in addition to conformal symmetry:
\begin{itemize}
\item \textbf{Existence of OPEs}. This leads to the fundamental property that $N$-point correlators reduce to $3$-point correlators.
 \item \textbf{Existence of a stress-energy tensor}. A CFT may be called \textit{local} if it has a stress-energy tensor, otherwise it is \textit{nonlocal}. This terminology can be confusing: the cluster connectivities of the $Q$-state Potts model are nonlocal observables on the lattice, but in the critical limit the model has a stress-energy tensor and no OPEs. Nonlocal CFT has been called \textit{conformal theory} by Schwarz \cite{sch15}, but this name is not widely used, even though the subject is quite active \cite{bbel26}.
 \item \textbf{Operator formalism and state-field correspondence}. Fields are operators acting on a space of states, moreover there is a bijection between states and fields. This axiom means little on the sphere, as we can reconstruct states and operators from correlation functions. This axiom becomes constraining if we require that the $0$-point correlator on $S^1\times S^{d-1}$ is a trace over the space of states. In 2d, this implies that the torus partition function is modular invariant, which is not the case for systems with large spectrums such as the set of all Coulomb gas integrals.
 \item \textbf{Existence on general spaces}. Can a CFT exist on the sphere only? In 2d, examples include
 submodels of minimal models \cite{bcm24}, and generalised minimal models \cite{rib24}.
 \item \textbf{Single-valuedness of correlators} may be relaxed in 2d, leading to fields with non-integer spins. Half-integer spins lead to \textit{fermionic CFT} \cite{rw20}, fractional spins to \textit{parafermionic CFT}.
 But as we saw in the case of free bosons in Section \ref{sec:fb}, multivaluedness raises issues with the definition of OPEs.
\end{itemize}
In our view, a CFT must have OPEs, a CFT needs not have a stress-energy tensor, and the rest is debatable.

\subsubsection{Lessons for critical loop models}

We have justified the observation that correlators are combinations of conformal blocks. While this observation was puzzling in the absence of OPEs, we have now derived it from the weaker axiom of correlator expansions. Furthermore, Proposition \ref{prop:shift} explains why conformal blocks combine into interchiral blocks. And we can now clarify the relations between various models: the critical $O(n)$ model, $Q$-state Potts model and $PSU(n)$ models are CFTs if $n\in\mathbb{N}_{\geq 2}$ or $Q\in\mathbb{N}_{\geq 2}$, they are conformal correlator systems for generic values or $n$ or $Q$, and for any given value of $\beta$ they are all subsystems of one larger conformal correlator system. Moreover, some of our technical results may be relevant for solving these models:
\begin{itemize}
 \item We knew that it was not enough to compute sphere $3$-point correlators, we now know that sphere $4$-point correlators are in principle sufficient.
 \item As we argued in Section \ref{sec:clm}, allowing complex spins is necessary for solving critical loop models in the analytic bootstrap approach.
 \item In fact, complex spins may lead to simplifications. In structure constants, fractional values of the second Kac index $s$ lead to complicated trigonometric coefficients. Structure constants may be simpler when written as functions of the complex variable $e^{i\pi s}$.
\end{itemize}

\subsubsection{Constructing conformal correlator systems}

Finally, while we have followed an axiomatic approach, we may wonder how to construct conformal correlator systems.
Which constructive approaches to CFT can be generalized to conformal correlator systems? The probabilistic constructions called Schramm--Loewner Evolutions and Conformal Loop Ensembles can describe non-local objects that may not admit OPEs, such as correlators related to the backbone exponent \cite{nqsz24}. Moreover, spiral SLEs have been proposed for describing fields with arbitrary spins, and they can likely be used for constructing their correlators \cite{hpw25}.

\section*{Acknowledgements}

\begin{itemize}
 \item I am grateful to Antoine Bourget, Philippe Di Francesco, Bertrand Eynard, Riccardo Guida, Quentin Lamouret, Eric Perlmutter, and Paul Roux, for stimulating discussions.
 \item I would like to thank Max Downing, Paul Roux and Dalimil Maz\'a\v{c} for comments and suggestions on the draft of this text.
 \item Artificial intelligence was not used in this work.
\end{itemize}

% Do not forget Julia notebook: revise, comment, send to arXiv.

%\pagebreak

\bibliographystyle{morder7}
\bibliography{992}

\end{document}